\documentclass[11pt,a4paper]{article}
\usepackage{amssymb}
\usepackage{graphicx}
\usepackage{amsmath}
\usepackage{makeidx}
\usepackage{indentfirst}
\usepackage[T1]{fontenc}
\usepackage[utf8]{inputenc}
\usepackage{authblk}

\newcounter{resultnum}[section]
\newcounter{conclusionnum}[section]
\newcounter{conditionnum}[section]
\newcounter{conjecturenum}[section]
\newcounter{examplenum}[section]
\newcounter{exercisenum}[section]
\newcounter{lemmanum}[section]
\newcounter{notationnum}[section]
\newcounter{theoremnum}[section]
\newcounter{definitionnum}[section]
\newcounter{corollarynum}[section]
\newcounter{remarknum}[section]
\newcounter{propositionnum}[section]
\newcounter{acknowledgementnum}[section]
\newcounter{algorithmnum}[section]
\newcounter{axiomnum}[section]
\newcounter{casenum}[section]
\newcounter{claimnum}[section]
\newcounter{summarynum}[section]
\newcounter{problemnum}[section]
\begin{document}

\title{\textbf{Effective Einstein Cosmological Spaces\\
for Non-Minimal Modified Gravity}}
\date{March 19, 2015}

\author[1]{Emilio Elizalde\thanks{elizalde@ieec.uab.es}}%

\author[2]{Sergiu I. Vacaru\thanks{sergiu.vacaru@uaic.ro}}%

\affil[1]{\small Instituto de Ciencias del Espacio, Consejo Superior de Investigaciones Cient\'{\i}ficas,
\newline ICE-CSIC and IEEC, Facultat de Ci\`encies, Campus UAB,  Torre C5-Parell-2a planta,
\newline 08193 Bellaterra (Barcelona) Spain%
\newline {\qquad }
}

\affil[2]{\small Theory Division, CERN, CH-1211, Geneva 23, Switzerland \footnote{associated visiting research affiliation};\ and
\newline
 Rector's Office, Alexandru Ioan Cuza University,\
 Alexandru Lapu\c sneanu street, \newline  nr. 14, UAIC - Corpus R, office 323;\
 Ia\c si,\ Romania, 700057
}%

\renewcommand\Authands{ and }

\maketitle

\begin{abstract}
Certain off-diagonal vacuum and nonvacuum configurations in Einstein gravity
can mimic physical effects of modified gravitational theories of $%
f(R,T,R_{\mu \nu}T^{\mu \nu})$ type. We prove this statement by constructing
exact and approximate solutions which encode certain models of covariant
Ho\v rava type gravity with dynamical Lorentz symmetry breaking.
Off-diagonal generalizations of de Sitter and nonolonomic $\Lambda$CDM
universes are constructed which are generated through nonlinear
gravitational polarization of fundamental physical constants and which model
interactions with non-constant exotic fluids and effective matter. The
problem of possible matter instability for such off-diagonal deformations in
(modified) gravity theories is discussed.

\vskip0.1cm


\vskip3pt

\end{abstract}







\section{Introduction}

There are various extensions of general relativity, GR, theory. \ Some of
the most popular are $f(R),f(R,T)$, and $f(\mathbf{R},\mathbf{T},F)$---which
we will here generically call $f$-modified theories, or modified gravity
theories, MGTs. \ For such modifications, the standard Lagrangian for GR,
namely as $\mathcal{L}=R$, on a pseudo-Riemannian manifold, $V$---where $R$
is the Ricci scalar curvature for the Levi-Civita connection, $\nabla $---is
modified by the addition of a functional, $f(R,...)$, of the Ricci scalar
only, in the first case, of $R$ and the torsion tensor, for a "boldface"
symbol $\mathbf{T}=\{T_{\beta \gamma }^{\alpha }\}$, the energy-momentum
tensor for matter, $T_{\beta \gamma },$ and/or its trace $T=T_{\alpha
}^{\alpha }$ (in the second), and of a generalized Ricci scalar $\mathbf{R}$%
, and a Finsler generating function, $F$, in the third case (such values may
be defined on the tangent bundle $TV$), etc. Classes of MGTs of these kinds
can be successfully constructed. Corresponding reconstruction procedures,
able to mimic the ${\Lambda }$CDM model including the dark energy epochs and
the transitions between the different main stages of the universe evolution
can be elaborated, \ see reviews of results in Refs.~\cite{revfmod}.

Several MGTs are actually related with different forms of the so-called
covariant Ho\v{r}ava gravity associated with a dynamical breaking of Lorentz
invariance \cite{covhl}, and with further developments, as well, including
generic off-diagonal solutions, Lagrange-Hamilton-Finsler like
generalizations, A-brane models, and gauge like gravity \cite{vhl}. Theories
of this kind can be constructed in a power-counting renormalizable form or
as nonholonomic brane configurations which correspond to power-law versions
of actions of type $f(R,T,R_{\mu \nu }T^{\mu \nu })$ \cite{odgom,voffds}. \
In general, such models are with commutative and/or noncommutative
parameters, and off-diagonal metrics for wapred/trapped solutions, Lorentz
violations, nonlinear dispersion relations and locally anisotropic
re-scaling, and effective polarizations of constants~\cite{finslmavr}.

Gravitational and matter field equations in GR and various MGTs usually
consist of very sophisticated systems of nonlinear partial differential
equations (PDEs) and request advanced numeric, analytic and geometric
techniques for constructing exact and approximate solutions. After a series
of assumptions of \textquotedblleft high symmetry\textquotedblright\ of the
relevant interactions (for spherical, cylindrical or torus ans\"{a}tze, with
a possible additional Lie group interior symmetry), such systems of
nonlinear PDEs are usually transformed into much more simplified systems of
nonlinear ordinary differential equations (ODEs). In such cases, some
classes of exact solutions can be obtained in explicit form being
paramterized by integration constants (see the monographs \cite{monexsol}
for reviews of some results in GR). At present, it is already possible to
elaborate a geometric techniques \cite{voffds}, the so-called anholonomic
frame deformation method (AFDM), for decoupling and generating solutions of
PDEs which to involve ansatz resulting in ODEs. We can construct more
general classes of exact solutions in MGTs and GR depending generically on
two, three and four variables in a \textquotedblleft less
symmetric\textquotedblright\ form for four-dimensional (4-d), and
extra-dimensional models using advanced geometric, analytic and numerical
methods.

The main goal of this article is to apply the AFDM for constructing of exact
off-diagonal solutions corresponding to cosmological models of MGTs of type $%
f(R,T,R_{\mu \nu }T^{\mu \nu })$ and to study the conditions under which
such configurations can be alternatively modeled as effective Einstein
spaces with nontrivial off-diagonal parametric vacuum and non-vacuum
configurations. There will be studied analogous FLRW cosmological dynamics
and a reconstruction procedure of the ${\Lambda }$CDM universe. We shall not
work with exotic anisotropic fluid configurations as in \cite{covhl,odgom},
but rather with off-diagonal deformations of de Sitter solutions \cite{vhl}.
The problem of matter instabilities in MGTs and GR will be analyzed
considering solutions describing nonholonomic cosmological configurations.

\section{Off-diagonal interactions in modified gravity and cosmology}

\label{s2}

\subsection{Off-diagonal metrics modelling dark energy}

In a spatially flat spacetime, we can consider diagonal quadratic form%
\begin{equation*}
ds^{2}=\mathring{g}_{\alpha }(t)(du^{\alpha })^{2}=\mathring{a}%
^{2}(t)[(dx^{1})^{2}+(dx^{2})^{2}+(dy^{3})^{2}]-dt^{2},
\end{equation*}%
for local coordinates $u^{\alpha }=(x^{i},y^{3},y^{4}=t),$ when $i=1,2.$ The
FLRW equations are $\frac{3}{\kappa ^{2}}\mathring{H}^{2}=\mathring{\rho}$
and $\mathring{\rho}^{\diamond }+3\mathring{H}(\mathring{\rho}+\mathring{p}%
)=0,$ where $\mathring{\rho}$ and $\mathring{p}$ are, respectively, the
total energy and pressure of a perfect fluid (pressureless or just
radiation), $\mathring{H}:=\mathring{a}^{\diamond }/\mathring{a}$ for $%
\mathring{a}^{\diamond }:=\partial \mathring{a}/\partial t=\partial _{4}%
\mathring{a}=\partial _{t}\mathring{a}.$ The constant $\kappa ^{2}$ is
related to the gravitational (Newton) constant.\footnote{%
We have to introduce a system of notations which is different from that in
standard cosmology. This will be convenient for constructing cosmological
models with generic off-diagonal metrics.} Various models have been studied
(see reviews and references in \cite{revfmod}), in order to explain the
observational data of an accelerating universe, and dark energy, DE, and
effective, real and/or exotic matter with an equation of state (EoS) of
phantom kind, $p=\varpi \rho ,$ with $\ \varpi <-1.$

The simplest model of phantom DE is given by $\ \frac{3}{\kappa ^{2}}%
H_{DE}^{2}=\rho _{DE}$ and $\rho _{DE}^{\diamond }+3H_{DE}(1+\varpi )\rho
_{DE}=0,$which for $\varpi <-1$ admits an exact solution
\begin{equation}
H_{DE}=\frac{2}{3(1+\varpi )(t_{s}-t)}.  \label{brip}
\end{equation}%
This solution has a finite-time future singularity (Big Rip) at $t=t_{s}.$

More general models have been elaborated for the Hubble function $H(t)$
determined by a phantom DE coupled with DM, via a coupling constant, $Q,$ \
and the conservation law%
\begin{equation*}
\rho _{DE}^{\diamond }+3H(1+\varpi )\rho _{DE}=-Q\rho _{DE},\ \rho
_{DM}^{\diamond }+3H\rho _{DM}=Q\rho _{DM}.
\end{equation*}%
The solutions are $\ \rho _{DE}=\ ^{0}\rho _{DE}\ e^{-3(1+\varpi )}e^{-Qt}$
and $\rho _{DM}a^{3}=Q\ ^{0}\rho _{DE}\int\nolimits^{t}dt^{\prime
}e^{-3\varpi }e^{-Qt},$ respectively, where $\ ^{0}\rho _{DE}$ is an
integration constant and the EoS is taken to be $p=\varpi \rho _{DE}.$ These
functions can be used for the second FLRW equation, $\ -\frac{1}{\kappa ^{2}}%
(2H^{\diamond }+3H^{2})=p.$ We have the solution
\begin{equation}
H=-Q/3(1+\varpi ),  \label{dssol}
\end{equation}
which corresponds to the de Sitter space evolution, $a(t)=a_{0}e^{-Qt/3(1+%
\varpi )},$ where $a_{0}$ is determined from $a_{0}^{3(1+\varpi )}=-\frac{%
3\kappa ^{2}}{Q^{2}}(1+\varpi )^{2}\varpi $ $\ ^{0}\rho _{DE}.$ The value of
$H$ in (\ref{dssol}) is positive for $\varpi <-1.$ This does not mean that
the Big Rip singularity in (\ref{brip}) can be avoided, but just shows that
the coupling of the phantom DE and DM gives a possibility that the universe
could evolve in the de Sitter phase. The first FLRW equation, $\frac{3}{%
\kappa ^{2}}H^{2}=\rho _{DE}+\rho _{DM}$, imposes the relation $\rho
_{DM}=(1+\varpi )\rho _{DE}.$ We can consider a de Sitter solution as an
attractor, with $\varpi \sim -4/3,$ i.e. $-(1+\varpi )\sim 1/3,$ which is
almost independent from the initial condition; we get a solution of the
so-called coincidence problem. If DE does not couple with DM, $\rho
_{DM}\sim a^{-3}$ and $\rho _{DE}\sim a^{-3(1+\varpi )}.$ The observed $1/3$
ratio of DE and DM is not satisfied which results in a coincidence problem.

Since the DE-DM coupling does not always remove the singularity and there is
no such fluid with constant EoS parameter, models were considered which are
proportional to a power of the scalar curvature, for instance, $%
p_{fluid}\propto R^{1+\epsilon },$ for $\epsilon >0,$ and the total EoS
parameter is greater than $-1.$ A Big Rip does not occur for large
curvature. Two variants of theories have been exploited where this kind of
inhomogeneous effective fluid matter is realized. A conformal anomaly and
other quantum effects or by some modified model of gravity, for instance,
when the gravitational Lagrange density $R\rightarrow f(R)=R+R^{\varkappa }.$
For $1<\varkappa <2$, we have that solutions with $\ Ht\sim -\frac{%
(\varkappa -1)(2\varkappa -1)}{\varkappa -2}$ and $w_{eff}\sim
-1-2H^{\diamond }/3H^{2}>-1$ do not result in a Big Rip or any other kind of
future singularity, see similar classical and quantum arguments as
motivations to study $f$-gravity \cite{revfmod,covhl,odgom,massgr}.

Various classes of off-diagonal solutions were studied which can be
constructed by geometric methods in MGTs \cite{voffds,vhl,voffdmgt}. Let us
briefly recall the main ideas supporting such an approach. Via frame
transforms, any metric can be parameterized by off-diagonal ans\"{a}tze for
metrics,
\begin{equation*}
g_{\underline{\alpha }\underline{\beta }}=\left[
\begin{array}{cccc}
g_{1}+\omega ^{2}(w_{1}^{\ 2}h_{3}+n_{1}^{\ 2}h_{4}) & \omega
^{2}(w_{1}w_{2}h_{3}+n_{1}n_{2}h_{4}) & \omega ^{2}w_{1}h_{3} & \omega
^{2}n_{1}h_{4} \\
\omega ^{2}(w_{1}w_{2}h_{3}+n_{1}n_{2}h_{4}) & g_{2}+\omega ^{2}(w_{2}^{\
2}h_{3}+n_{2}^{\ 2}h_{4}) & \omega ^{2}w_{2}h_{3} & \omega ^{2}n_{2}h_{4} \\
\omega ^{2}w_{1}h_{3} & \omega ^{2}w_{2}h_{3} & \omega ^{2}h_{3} & 0 \\
\omega ^{2}n_{1}h_{4} & \omega ^{2}n_{2}h_{4} & 0 & \omega ^{2}h_{4}%
\end{array}%
\right] ,  \label{ans1a}
\end{equation*}%
where the coefficients are functions of type: $g_{1}=g_{2}\sim e^{\psi
(x^{i})}$ and $n_{i}(x^{k})$ (we can fix certain constants for corresponding
classes of generating, $\Phi (x^{k},t),$ and integration functions), $%
h_{a}[\Phi (x^{k},t)]\sim h_{a}(t),$ [for $a=3,4],w_{i}[\Phi (x^{k},t)]\sim
w_{i}(t)$ and $\omega (x^{k},t)\sim \omega (t),$ were found to generate
exact (in general, nonhomogeneous) cosmological solutions in modified
gravity theories. Such generic off-diagonal metrics\footnote{%
Which cannot be diagonalized by coordinate transformations.} can be
represented as
\begin{equation}
ds^{2}=a^{2}(t)[(e^{1})^{2}+(e^{2})^{2}]+a^{2}(t)\widehat{h}_{3}(t)(\widehat{%
\mathbf{e}}^{3})^{2}+(\widehat{\mathbf{e}}^{4})^{2},  \label{ans1b}
\end{equation}%
with respect to N-adapted frames $\ \widehat{\mathbf{e}}%
^{3}=dy^{3}+n_{i}dx^{i}$ and $\widehat{\mathbf{e}}^{4}=dt+w_{i}(t)dx^{i}.$

We can formulate certain well-defined conditions, see Sect.~\ref{safdm},
when off-diagonal deformations $\mathring{g}_{\alpha }(t)\rightarrow g_{%
\underline{\alpha }\underline{\beta }}(x^{k},t)$ $\sim g_{\underline{\alpha }%
\underline{\beta }}(t)$ define new classes of cosmological models. Such
deformations mimic contributions from $f$--gravity encoded into the data for
$\omega (t),w_{i}(t)$ etc. when corresponding formulas are nonlinear
functionals relating generating functions to the (effective) matter sources.
The off-diagonal configurations are equivalently modeled as solutions of
some effective field equations $\mathbf{\check{R}}_{\ \beta }^{\alpha }=%
\check{\Lambda}\delta _{\ \beta }^{\alpha }.$ In result, various classes of
cosmological solutions in MGTs can be alternatively modeled by metrics of
type (\ref{ans1b}). In all cases, the scaling factor $a(t)$ is nonlinearly
determined by the coefficients $w_{i}(t)$ and $h_{a}(t)$ via a generating
function $\Phi (t)$ and an effective source $\Upsilon (t).$ It is possible
to model ${\Lambda }$CDM cosmology and analogously DE and DM effects with $%
\rho _{DE}+\rho _{DM}$ encoded into $\Phi (t)$ and $\Upsilon (t),$ but with
respect to the adapted frames $\widehat{\mathbf{e}}^{a}(t).$ Solutions with
small off-diagonal deformations of metrics may be interpreted in accordance
with observational data if the factor $a(t)$ is chosen to determine
nonlinearly/ parameterically, for instance, an effective $H(t)$\ (\ref{dssol}%
) with cosmological evolution from a spacetime background encoding $f(R)$%
-modifications.

Let us suppose that we have found a cosmological solution (\ref{ans1b}) in a
given MGT and analyze how this metric can be formally diagonalized for
deformations with a small real parameter $\varepsilon $ (when $0\leq
\varepsilon $ $\ll 1).$ We can consider \textquotedblleft
homogeneous\textquotedblright\ approximations of type $\widehat{h}%
_{3}(t)\approx 1+\varepsilon \widehat{\chi }_{3}(t),w_{i}(t)\sim \varepsilon
\check{w}_{i}(t)$ and $n_{i}\sim \varepsilon \check{n}_{i}.$ $\ $On
inhomogeneity effects in cosmology and possible physical models, see \cite%
{ellisgfr}. In a more general context, it is possible \ to elaborate on
\textquotedblleft small\textquotedblright\ local anisotropic deformations
depending on space like coordinates when $\widehat{\chi }%
_{3}(x^{k},t),w_{i}(t)\sim \varepsilon \check{w}_{i}(x^{k},t)$ and $%
n_{i}\sim \varepsilon \check{n}_{i}(x^{k}).$ Some amount of anisotropy is
compatible with observational data in various gravity and cosmological
theories (see \cite{goedel,krasinski}, for reviews of various approaches
related to GR and generalizations of Bianchi, Kasner and G\"{o}del type
configurations; \cite{voffdmgt}, for off--diagonal configurations; and \cite%
{appleby}, for $f$--gravity theories). We note also that the approximation $%
\widehat{h}_{3}(t)\approx 1+\varepsilon \widehat{\chi }_{3}(t)$ can be very
restrictive---one can consider more general classes of solutions with
arbitrary $\widehat{h}_{3}(t).$

The metrics with small off--diagonal deformations on $\varepsilon $ and
rescaling $\mathring{a}(t)\rightarrow a(t)$, can be written as
\begin{equation}
ds^{2}=a^{2}(t)[(e^{1})^{2}+(e^{2})^{2}]+a^{2}(t)[1+\varepsilon \widehat{%
\chi }_{3}(t)](dy^{3}+\varepsilon \check{n}_{i}dx^{i})^{2}+(dt+\varepsilon
\check{w}_{i}(t)dx^{i})^{2}.  \label{ans1bb}
\end{equation}%
Below, we shall discuss how it is possible to construct subclasses of
off--diagonal configurations in a $\widehat{\mathbf{f}}(\widehat{\mathbf{R}},%
\mathbf{...})$ gravity where $\ \widehat{\mathbf{\Upsilon }}$
) goes into $\check{\Lambda}$   and $\check{\Phi}^{2}=\check{%
\Lambda}^{-1}[\widehat{\Phi }^{2}|\ \widehat{\mathbf{\Upsilon }}|+\int
d\zeta \ \widehat{\Phi }^{2}\partial _{\zeta }|\ \widehat{\mathbf{\Upsilon }}%
|]$  results in $\widehat{\mathbf{f}}\rightarrow \mathbf{\check{%
f}=\check{R},}$ with an effective $\mathbf{\check{R}}_{\ \beta }^{\alpha }=%
\check{\Lambda}\delta _{\ \beta }^{\alpha }.$ We will be able to reproduce
the $\Lambda $CDM model provided the metric (\ref{ans1bb}) defines certain
classes of solutions constructed for a corresponding effective action in GR,
namely
\begin{equation}
S=\frac{1}{\kappa ^{2}}\int \delta ^{4}u\sqrt{|\ ^{\varepsilon }\mathbf{g}%
_{\alpha \beta }|}(\ ^{\varepsilon }\mathbf{\check{R}}-2\check{\Lambda}+\
_{m}\mathcal{L}(\ ^{\varepsilon }\mathbf{g}_{\alpha \beta },\ _{m}\Psi )).
\label{actex}
\end{equation}%
The Ricci scalar $\ ^{\varepsilon }\mathbf{\check{R}=\check{R}(}%
a,\varepsilon )$ is constructed for $\ ^{\varepsilon }\mathbf{g}_{\alpha
\beta }$ with coefficients of (\ref{ans1bb}), $\check{\Lambda}$ is an
effective cosmological constant used for nonholonomic deformations, and $\
_{m}\mathcal{L}$ is considered for certain effective matter fields with
certain pressure $\ _{m}p$ and energy--density $\ _{m}\rho .$ The EoS are
chosen, for simplicity, to correspond to an effective de Sitter
configuration determined by $\check{\Lambda},$ where $\check{\varpi}:=\check{%
p}_{\Lambda }/\check{\rho}_{\Lambda }=-1$, with pressure $\check{p}_{\Lambda
}$ and energy--density $\check{\rho}_{\Lambda }.$

We can describe the theories determined by (\ref{actex}) and (\ref{ans1bb})
with respect to nonholonomic (non-integrable) dual frames $\widehat{\mathbf{e%
}}^{\alpha }=(e^{i},\widehat{\mathbf{e}}^{a}),$ which is convenient for
constructing off--diagonal solutions, or to redefine the constructions with
respect to local coordinate coframes $du^{\alpha }=(dx^{i},dy^{a}),$ where
certain analog of the FLRW metric and $\Lambda $CDM like theories can be
analyzed. For $\varepsilon \rightarrow 0,$ the metric (\ref{ans1bb})
transforms into
\begin{equation}
ds^{2}=a^{2}(t)[(e^{1})^{2}+(e^{2})^{2}+(dy^{3})]^{2}+dt^{2},  \label{ans1bc}
\end{equation}%
which is just the FLRW metric but with a re--scaled factor because of $%
\widehat{\mathbf{\Upsilon }}$ $\rightarrow $ $\check{\Lambda}$ and $\widehat{%
\Phi }\rightarrow \check{\Phi}.$

The corresponding effective Einstein equations with respect to the
nonholonomic frames are%
\begin{equation}
3H^{2}=\kappa ^{2}\ _{m}\rho +\check{\Lambda},2H^{\diamond }=-\kappa ^{2}(\
_{m}\rho +\ _{m}P+\check{\Lambda}),  \label{enst1c1}
\end{equation}%
where $H^{\diamond }:=a^{\diamond }/a.$ We can express $\ ^{\varepsilon }%
\mathbf{\check{R}}+\ _{m}\mathcal{L}\mathbf{=\ }^{a}\mathbf{\check{R}+}\
_{m}^{0}\mathcal{L+}\varepsilon \ _{m}^{1}\mathcal{L}$, where $\mathbf{\ }%
^{a}\mathbf{\check{R}}$ and $\ _{m}^{0}\mathcal{L}$ are computed for the
metric (\ref{ans1bc}) and $\ _{m}^{1}\mathcal{L}$ include all $\varepsilon $%
--deformations in (\ref{actex}). The term $\ _{m}^{1}\mathcal{L}$ results in
the effective splitting $_{m}\rho =\ _{m}^{0}\rho +\varepsilon \
_{m}^{1}\rho $ and $_{m}p=\ _{m}^{0}p+\varepsilon \ _{m}^{1}p.$ In this way,
we can encode the off--diagonal components as certain additional terms into
the matter source, or either consider them as a polarization of the
effective cosmological constant $\Lambda :=\check{\Lambda}+\varepsilon \ ^{1}%
\check{\Lambda}.$ We do not provide explicit formulas for the corrections
proportional to $\varepsilon $ because, in the end, we shall take smooth
limits $\varepsilon \rightarrow 0.$ The main constructions for nonholonomic
off--diagonal transforms are based on rescaling $\mathring{a}(t)\rightarrow
a^{2}(t)$ $\ $generated by the solutions with $\widehat{\mathbf{\Upsilon }}$
$\rightarrow $ $\check{\Lambda}$ and $\widehat{\Phi }\rightarrow \check{\Phi}%
.$ Possible small inhomogeneous and locally anisotropic contributions, and
concordance with observational data, can be estimated similarly to those
presented, e.g., in \cite{appleby}. In coordinate frames, Eqs. (\ref{enst1c1}%
) are written as $3H^{2}=\kappa ^{2}\ _{m}^{0}\rho +\Lambda $, and $\
2H^{\diamond }=-\kappa ^{2}(\ _{m}^{0}\rho +\ _{m}^{0}P+\Lambda )$. For $%
\varepsilon \rightarrow 0,$ the diagonalized solutions are determined by $a$
(and not by $\mathring{a}$) and can be parameterized to define and effective
$\Lambda $CDM like model where $a=a_{c}e^{H_{c}t},$ for a positive constant $%
a_{c}.$ Thus, MGTs with equivalent off--diagonal encodings of $f$--gravity
seem to result in realistic cosmological models, at least for small
parametric $\varepsilon $--deformations.

We conclude that within certain assumptions, various possible $f(R)$%
--nonlinear modifications can be encoded into off--diagonal terms and some
effective $a(t),$ $\widehat{h}_{3}(t),w_{i}(t)$ via nonlinear interactions.
This can be done for more general classes of cosmological solutions with
nonlinear gravitational interactions restructuring the spacetime aether
before considering certain small $\varepsilon $--parameters.

\subsection{Geometric preliminaries}

We consider a pseudo-Riemannian manifold $V,$\ $\dim V=n+m,$ ($n,m\geq 2$).
A Whitney sum $\mathbf{N}$ is defined for its tangent space $TV,$ when $%
\mathbf{N}:\ TV=hTV\oplus vTV.$ This states a nonholonomic (equivalently,
non-integrable, or anholonomic) horizontal (h) and vertical (v) splitting,
or a nonlinear connection (\textit{N-connection}) structure. In local form,
it is determined by its coefficients $\mathbf{N}=\{N_{i}^{a}(u)\},$ when $%
\mathbf{N}=N_{i}^{a}(x,y)dx^{i}\otimes \partial /\partial y^{a}$ for certain
local coordinates $u=(x,y),$ or $u^{\alpha }=(x^{i},y^{a}),$ and $h$-indices
$i,j,...=1,2,...n$ and $v$-indices $a,b,...=n+1,n+2,...,n+m.$\footnote{%
The Einstein rule on index summation will be applied if the contrary is not
stated. Boldface letters are used in order to emphasize that an N-connection
spitting is considered on a spacetime manifold $\mathbf{V=(}V,\mathbf{N).}$}
Such a h-v-decomposition can be naturally associated with some N-adapted
frame or, respectively, dual frame structures, $\mathbf{e}_{\nu }=(\mathbf{e}%
_{i},e_{a})$ and $\mathbf{e}^{\mu }=(e^{i},\mathbf{e}^{a}),$
\begin{equation}
\mathbf{e}_{i}=\partial /\partial x^{i}-\ N_{i}^{a}(u)\partial /\partial
y^{a},\ e_{a}=\partial _{a}=\partial /\partial y^{a},\mbox{ and  }%
e^{i}=dx^{i},\ \mathbf{e}^{a}=dy^{a}+\ N_{i}^{a}(u)dx^{i}.  \label{nadif}
\end{equation}%
The nonholonomy relations hold $\ [\mathbf{e}_{\alpha },\mathbf{e}_{\beta }]=%
\mathbf{e}_{\alpha }\mathbf{e}_{\beta }-\mathbf{e}_{\beta }\mathbf{e}%
_{\alpha }=W_{\alpha \beta }^{\gamma }\mathbf{e}_{\gamma },$ with nontrivial
anholonomy coefficients $W_{ia}^{b}=\partial _{a}N_{i}^{b},W_{ji}^{a}=\Omega
_{ij}^{a}=\mathbf{e}_{j}\left( N_{i}^{a}\right) -\mathbf{e}_{i}(N_{j}^{a})$.
The coefficients $\Omega _{ij}^{a}$ define the N-connection curvature.

Any metric structure $\mathbf{g}$ on $\mathbf{V}$ (for physical
applications, we consider the signature $\left( +,+,+,-\right) $) can be
written in two equivalent ways: 1) with respect to a dual local coordinate
basis,
\begin{equation}
\mathbf{g}=\underline{g}_{\alpha \beta }du^{\alpha }\otimes du^{\beta },
\label{m1}
\end{equation}%
where $\ \underline{g}_{\alpha \beta }=\left[
\begin{array}{cc}
g_{ij}+N_{i}^{a}N_{j}^{b}g_{ab} & N_{j}^{e}g_{ae} \\
N_{i}^{e}g_{be} & g_{ab}%
\end{array}%
\right] ,$ or 2) as a distinguished metric (in brief, \textit{d-metric},
i.e. in N-adapted form,
\begin{equation}
\mathbf{g}=g_{\alpha \beta }(u)\mathbf{e}^{\alpha }\otimes \mathbf{e}^{\beta
}=g_{i}(x^{k})dx^{i}\otimes dx^{i}+g_{a}(x^{k},y^{b})\mathbf{e}^{a}\otimes
\mathbf{e}^{a}.  \label{dm1}
\end{equation}%

A linear connection is called distinguished,\textit{\ d-connection,} $%
\mathbf{D}=(hD,vD),$ if it preserves under parallelism a prescribed
N-connection splitting. Any $\mathbf{D}$ defines an operator of covariant
derivation, $\mathbf{D}_{\mathbf{X}}\mathbf{Y}$, for a d-vector field $%
\mathbf{Y}$ in the direction of a d-vector $\mathbf{X}.$ We note that any
vector $Y(u)\in T\mathbf{V}$ can be parameterized as a d-vector, $\mathbf{Y}%
= $ $\mathbf{Y}^{\alpha }\mathbf{e}_{\alpha }=\mathbf{Y}^{i}\mathbf{e}_{i}+%
\mathbf{Y}^{a}e_{a},$ or $\mathbf{Y}=(hY,vY),$ with $hY=\{\mathbf{Y}^{i}\}$
and $vY=\{\mathbf{Y}^{a}\},$ where the N-adapted base vectors and duals, or
covectors, are chosen in N-adapted form (\ref{nadif}). The local
coefficients of $\mathbf{D}_{\mathbf{X}}\mathbf{Y}$ can be computed for $%
\mathbf{D}=\{\mathbf{\Gamma }_{\ \alpha \beta }^{\gamma
}=(L_{jk}^{i},L_{bk}^{a},C_{jc}^{i},C_{bc}^{a})\}$ and h-v-components of $%
\mathbf{D}_{\mathbf{e}_{\alpha }}\mathbf{e}_{\beta }:=$ $\mathbf{D}_{\alpha }%
\mathbf{e}_{\beta }$ using $\mathbf{X}=\mathbf{e}_{\alpha }$ and $\mathbf{Y}=%
\mathbf{e}_{\beta }.$ There are used the terms d-vector, d-tensor, etc. for
any vector, tensor valued with coefficients defined in a N-adapted form with
respect to the necessary types of tensor products of N--elongate bases and
necessary $h$-$v$-decompositions. We can define three fundamental geometric
objects: the d-torsion, $\mathcal{T},$ the non-metricity, $\mathcal{Q},$ and
the d-curvature, $\mathcal{R},$ respectively defined by
\begin{equation}
\mathcal{T}(\mathbf{X,Y}):=\mathbf{D}_{\mathbf{X}}\mathbf{Y}-\mathbf{D}_{%
\mathbf{Y}}\mathbf{X}-[\mathbf{X,Y}],\mathcal{Q}(\mathbf{X}):=\mathbf{D}_{%
\mathbf{X}}\mathbf{g,\ }\mathcal{R}(\mathbf{X,Y}):=\mathbf{D}_{\mathbf{X}}%
\mathbf{D}_{\mathbf{Y}}-\mathbf{D}_{\mathbf{Y}}\mathbf{D}_{\mathbf{X}}-%
\mathbf{D}_{\mathbf{[X,Y]}}.  \notag
\end{equation}%
The N-adapted coefficients, $\ \mathcal{T}=\{\mathbf{T}_{\ \alpha \beta
}^{\gamma }=\left( T_{\ jk}^{i},T_{\ ja}^{i},T_{\ ji}^{a},T_{\ bi}^{a},T_{\
bc}^{a}\right) \},\mathcal{Q}=\mathbf{\{Q}_{\ \alpha \beta }^{\gamma }\}$
and \newline
$\mathcal{R}=\mathbf{\{R}_{\ \beta \gamma \delta }^{\alpha }\mathbf{=}\left(
R_{\ hjk}^{i},R_{\ bjk}^{a},R_{\ hja}^{i},R_{\ bja}^{c},R_{\ hba}^{i},R_{\
bea}^{c}\right) \},$ of such fundamental geometric objects are computed by
introducing $\mathbf{X}=\mathbf{e}_{\alpha }$ and $\mathbf{Y}=\mathbf{e}%
_{\beta },$ and $\mathbf{D}=\{\mathbf{\Gamma }_{\ \alpha \beta }^{\gamma }\}$
into the formulas above (see \cite{voffds} for details).

A d-connection $\mathbf{D}$ is compatible with a d-metric $\mathbf{g}$ if
and only if $\mathcal{Q}=\mathbf{Dg}=0.$ Any metric structure $\mathbf{g}$
on $\mathbf{V}$ is characterized by a unique metric compatible and
torsionless linear connection called the Levi-Civita (LC) connection, $%
\nabla .$ It should be noted that $\nabla $ is not a d-connection because it
does not preserve under parallelism the N-connection splitting.
Nevertheless, such a $h$-$v$ decomposition allows us to define N-adapted
distortions of any d-connection $\mathbf{D,}$
\begin{equation}
\mathbf{D}=\nabla +\mathbf{Z},  \label{distr}
\end{equation}%
with respective conventional ``non-boldface'' and ``boldface'' symbols for
the coefficients: $\nabla =\{\Gamma _{\ \beta \gamma }^{\alpha }\}$ and, for
the distortion d-tensor, $\mathbf{Z}=\{\mathbf{Z}_{\ \beta \gamma }^{\alpha
}\}. $

This stands for any prescribed $\mathbf{N}$ and $\mathbf{g}=h\mathbf{g}+v%
\mathbf{g,}$ but alternatively to $\nabla $, on $\mathbf{V}$, we can work
with the so-called canonical d-connection, $\widehat{\mathbf{D}},$ when
\begin{equation*}
(\mathbf{g,N})\rightarrow
\begin{array}{cc}
\mathbf{\nabla :} & \mathbf{\nabla g}=0;\ ^{\nabla }\mathcal{T}=0; \\
\widehat{\mathbf{D}}: & \widehat{\mathbf{D}}\mathbf{g}=0;\ h\widehat{%
\mathcal{T}}=0,v\widehat{\mathcal{T}}=0,hv\widehat{\mathcal{T}}\neq 0;%
\end{array}%
\end{equation*}%
are completely defined \textit{by the same} metric structure. The canonical
distortion d-tensor $\widehat{\mathbf{Z}}$ in the distortion relation of
type (\ref{distr}), $\widehat{\mathbf{D}}=\nabla +\widehat{\mathbf{Z}},$ is
an algebraic combination of the coefficients of the corresponding torsion
d-tensor $\widehat{\mathcal{T}}=\{\widehat{\mathbf{T}}_{\ \beta \gamma
}^{\alpha }\}.$ The respective coefficients of the torsions, $\widehat{%
\mathcal{T}}$ and $\ ^{\nabla }\mathcal{T}=0,$ and curvatures, $\widehat{%
\mathcal{R}}=\{\widehat{\mathbf{R}}_{\ \beta \gamma \delta }^{\alpha }\}$
and $\ ^{\nabla }\mathcal{R}=\{R_{\ \beta \gamma \delta }^{\alpha }\},$ of $%
\widehat{\mathbf{D}}$ and $\nabla $ can be defined and computed using
standard formulas. The coefficients $\widehat{\mathbf{T}}_{\ \beta \gamma
}^{\alpha }$ are not trivial but nonholonomically induced by anholonomy
coefficients $W_{\alpha \beta }^{\gamma }$ and certain off-diagonal
coefficients of the metric.

The Ricci tensors of $\widehat{\mathbf{D}}$ and $\nabla $ are computed in
the standard form, $\widehat{\mathcal{R}}ic=\{\widehat{\mathbf{R}}_{\ \beta
\gamma }:=\widehat{\mathbf{R}}_{\ \alpha \beta \gamma }^{\gamma }\}$ and $%
Ric=\{R_{\ \beta \gamma }:=R_{\ \alpha \beta \gamma }^{\gamma }\}.$ With
respect to N-adapted coframes (\ref{nadif}), the Ricci d-tensor $\widehat{%
\mathcal{R}}ic$ is characterized by four $h$-$v$ N-adapted coefficients {\
\begin{equation}
\widehat{\mathbf{R}}_{\alpha \beta }=\{\widehat{R}_{ij}:=\widehat{R}_{\
ijk}^{k},\ \widehat{R}_{ia}:=-\widehat{R}_{\ ika}^{k},\ \widehat{R}_{ai}:=%
\widehat{R}_{\ aib}^{b},\ \widehat{R}_{ab}:=\widehat{R}_{\ abc}^{c}\},
\label{driccic}
\end{equation}%
} and (an alternative to $\ R:=\mathbf{g}^{\alpha \beta }R_{\alpha \beta })$
scalar curvature, $\widehat{\mathbf{R}}:=\mathbf{g}^{\alpha \beta }\widehat{%
\mathbf{R}}_{\alpha \beta }=g^{ij}\widehat{R}_{ij}+g^{ab}\widehat{R}_{ab}.$

We emphasize that any (pseudo) Riemannian geometry can be equivalently
described by both geometric data $\left( \mathbf{g,\nabla }\right) $ and $(%
\mathbf{g,N,}\widehat{\mathbf{D}}).$ For instance, there are canonical
distortion relations $\widehat{\mathcal{R}}=\ ^{\nabla }\mathcal{R+}\
^{\nabla }\mathcal{Z}$ and $\widehat{\mathcal{R}}ic=Ric+\widehat{\mathcal{Z}}%
ic$, where the respective distortion d-tensors $\ ^{\nabla }\mathcal{Z}$ and
$\widehat{\mathcal{Z}}ic$ are computed by introducing $\widehat{\mathbf{D}}%
=\nabla +\widehat{\mathbf{Z}}$ into the corresponding formulas for curvature
and (\ref{driccic}). The canonical data $(\mathbf{g,N,}\widehat{\mathbf{D}})
$ provide an example of nonholonomic (pseudo-) Riemannian manifold which is
a standard one but enabled with a nonholonomic distribution determined by $(%
\mathbf{g,N}).$ If the coefficients $\Omega _{ij}^{a}=0$, such a
distribution is holonomic, i.e. integrable. Nevertheless, physical theories
formulated in terms of data as $\left( \mathbf{g,\nabla }\right) $, or $(%
\mathbf{g,N,}\widehat{\mathbf{D}})$, are not equivalent if certain
additional conditions are not imposed.

We can introduce the Einstein d-tensor of $\widehat{\mathbf{D}},$  $\widehat{%
\mathbf{E}}_{\alpha \beta }:=\widehat{\mathbf{R}}_{\alpha \beta }-\frac{1}{2}%
\mathbf{g}_{\alpha \beta }\ \widehat{\mathbf{R}}$, and construct a N-adapted
energy momentum tensor for a Lagrange density $\ ^{m}\mathcal{L}$ of the
matter fields, $\widehat{\mathbf{T}}_{\alpha \beta }:=-\frac{2}{\sqrt{|%
\mathbf{g}_{\mu \nu }|}}\frac{\delta (\sqrt{|\mathbf{g}_{\mu \nu }|}\ \ ^{m}%
\widehat{\mathcal{L}})}{\delta \mathbf{g}^{\alpha \beta }},~$performing a
N-adapted variational calculus with respect to frames (\ref{nadif}), and
considering that $\widehat{\mathbf{D}}$ is used as covariant derivative
instead of $\nabla .$ A nonholonomic deformation of Einstein's gravity is
constructed, being $\nabla \rightarrow $ $\widehat{\mathbf{D}}=\nabla +%
\widehat{\mathbf{Z}},$ with gravitational field equations
\begin{equation}
\widehat{\mathbf{R}}_{\alpha \beta }=\kappa ^{2}(\widehat{\mathbf{T}}%
_{\alpha \beta }-\frac{1}{2}\mathbf{g}_{\alpha \beta }\widehat{\mathbf{T}})
\label{nheeq}
\end{equation}%
for a conventional gravitational constant $\kappa ^{2}$ and $\widehat{%
\mathbf{T}}:=\mathbf{g}^{\mu \nu }\widehat{\mathbf{T}}_{\mu \nu }.$ Such
equations are different from the standard Einstein equations in GR because,
in general, $\widehat{\mathbf{R}}_{\alpha \beta }\neq R_{\alpha \beta }$ and
$\widehat{\mathbf{T}}_{\alpha \beta }\neq T_{\alpha \beta },$ where $%
T_{\alpha \beta }:=-\frac{2}{\sqrt{|\mathbf{g}_{\mu \nu }|}}\frac{\delta (%
\sqrt{|\mathbf{g}_{\mu \nu }|}\ \ ^{m}\mathcal{L})}{\delta \mathbf{g}%
^{\alpha \beta }}$ for $\ ^{m}\mathcal{L}[\mathbf{g}_{\alpha \beta },\nabla ]%
\mathcal{\neq \ }^{m}\widehat{\mathcal{L}}[\mathbf{g}_{\alpha \beta },%
\widehat{\mathbf{D}}].$

LC-configurations can be extracted from certain classes of solutions of
Eqs.~(\ref{nheeq}) if additional conditions are imposed, resulting in zero
values for the canonical d-torsion, $\widehat{\mathcal{T}}=0$. In N-adapted
coefficient form, such conditions are equivalent to
\begin{equation}
\widehat{T}_{\ jk}^{i}=\widehat{L}_{jk}^{i}-\widehat{L}_{kj}^{i},\widehat{T}%
_{\ ja}^{i}=\widehat{C}_{jb}^{i},\widehat{T}_{\ ji}^{a}=-\Omega _{\ ji}^{a},%
\widehat{T}_{aj}^{c}=\widehat{L}_{aj}^{c}-e_{a}(N_{j}^{c}),\widehat{T}_{\
bc}^{a}=\ \widehat{C}_{bc}^{a}-\ \widehat{C}_{cb}^{a}.  \label{dtors}
\end{equation}
It should be emphasized that we are able to find generic off-diagonal
solutions of the Einstein equations in GR depending on three and more
coordinates for $\widehat{\mathbf{D}}\rightarrow \nabla ,$ when $\widehat{%
\mathbf{R}}_{\alpha \beta }\rightarrow $ $R_{\alpha \beta }$ and $\widehat{%
\mathbf{T}}_{\alpha \beta }\rightarrow $ $T_{\alpha \beta },$ if the
nonholonomic constraints (\ref{dtors}) are imposed after certain classes of
solutions were found for $\widehat{\mathbf{D}}\neq \nabla .$ But we are not
able to decouple such systems of nonlinear PDEs if the zero torsion
condition for $\nabla $ is imposed from the very beginning.

\subsection{Nonholonomic $f$-modified gravity theories}

Different models of modified gravity are formulated for independent metric
and linear connection fields with a corresponding Palatini type variational
formulation (see \cite{revfmod}). The gravitational and matter field
equations in MGTs consist in very sophisticate systems of nonlinear PDEs for
which finding exact solutions is a very difficult technical task, even for
the simplest diagonal ans\"{a}tze with the coefficients of the metrics and
connections depending on just one (time or space variable). Nevertheless,
the AFDM \cite{voffds} seems to work efficiently and allows to construct
off-diagonal solutions in MGTs and GR \cite{voffdmgt}.

Let us consider three classes of equivalent MGTs defined for the same metric
field $\mathbf{g}=\{g_{\mu \nu }\}$ but with different actions (and related
functionals) for gravity, $\ ^{g}S,$ and matter, $\ ^{m}S,$ fields,%
\begin{eqnarray}
\mathcal{S} &=&\ ^{g}\mathcal{S}+\ ^{m}\mathcal{S}=\frac{1}{2\kappa ^{2}}%
\int f(R,T,R_{\alpha \beta }T^{\alpha \beta })\sqrt{|g|}d^{4}u+\int \ ^{m}%
\mathcal{L}\sqrt{|g|}d^{4}u  \notag \\
&=&\ ^{g}\widehat{\mathbf{S}}+\ ^{m}\widehat{\mathbf{S}}=\frac{1}{2\kappa
^{2}}\int \widehat{\mathbf{f}}(\widehat{\mathbf{R}},\widehat{\mathbf{T}},%
\widehat{\mathbf{R}}_{\alpha \beta }\widehat{\mathbf{T}}^{\alpha \beta })%
\sqrt{|\widehat{\mathbf{g}}|}\mathbf{d}^{4}u+\int \ ^{m}\widehat{\mathbf{L}}%
\sqrt{|\widehat{\mathbf{g}}|}\mathbf{d}^{4}u  \notag \\
&=&\ ^{g}\mathbf{\check{S}}+\ ^{m}\mathbf{\check{S}}=\frac{1}{2\kappa ^{2}}%
\int \mathbf{\check{R}}\sqrt{|\mathbf{\check{g}}|}\mathbf{d}^{4}u+\check{%
\Lambda}\int \sqrt{|\mathbf{\check{g}}|}\mathbf{d}^{4}u.  \label{mgts}
\end{eqnarray}%
We use boldface $\mathbf{d}^{4}u$ in order to emphasize that the integration
volume is for N-elongated differentials (\ref{nadif}), $\kappa ^{2}$ is the
gravitational coupling constant, the values with \textquotedblleft $\symbol{%
94}$\textquotedblright\ are computed for a canonical d-connection $\widehat{%
\mathbf{D}}$ and the values with \textquotedblleft $\vee $%
\textquotedblright\ for re-defined geometric data $(\mathbf{\check{g},\check{%
N},\check{D}})$ for certain nonholonomic frame transforms and nonholonomic
deformations $g_{\alpha \beta }\sim \widehat{\mathbf{g}}_{\alpha \beta }\sim
\mathbf{\check{g}}_{\alpha \beta }.$\footnote{%
see details in Sects.~\ref{sslc} and \ref{sseflrw}.} For simplicity, we
consider matter actions which only depend on the coefficients of a metric
field and not on their derivatives, $\widehat{\mathbf{T}}^{\alpha \beta }=\
^{m}\widehat{\mathbf{L}}\ \widehat{\mathbf{g}}^{\alpha \beta }+2\delta (\
^{m}\widehat{\mathbf{L}})/\delta \widehat{\mathbf{g}}_{\alpha \beta }.$

We assume that the matter content of the universe can be approximated by a
perfect fluid,
\begin{equation}
\widehat{\mathbf{T}}_{\alpha \beta }=p\widehat{\mathbf{g}}_{\alpha \beta
}+(\rho +p)\widehat{\mathbf{v}}_{\alpha }\widehat{\mathbf{v}}_{\beta }
\label{dsourc}
\end{equation}%
is defined for certain (effective) energy and pressure densities,
respectively, $\widehat{\mathbf{v}}_{\alpha }$ being the four-velocity of
the fluid for which $\widehat{\mathbf{v}}_{\alpha }\widehat{\mathbf{v}}%
^{\alpha }=-1$ and $\widehat{\mathbf{v}}^{\alpha }=(0,0,0,1)$ in N-adapted
comoving frames/coordinates. Frame transforms of metrics of type $\widehat{%
\mathbf{g}}_{\alpha \beta }=\mathbf{e}_{\ \alpha }^{\alpha ^{\prime }}%
\mathbf{e}_{\ \beta }^{\beta ^{\prime }}\mathring{g}_{\alpha ^{\prime }\beta
^{\prime }},$ will be studied beginning with the FLRW diagonalized element
\begin{equation}
d\mathring{s}^{2}=\mathring{g}_{\alpha ^{\prime }\beta ^{\prime }}du^{\alpha
^{\prime }}du^{\beta ^{\prime }}=\mathring{a}^{2}(t)[dr^{2}+r^{2}d\theta
^{2}+r^{2}\sin ^{2}\theta d\varphi ^{2}]-dt^{2},=\mathring{a}%
^{2}(t)[dx^{2}+dy^{2}+dz^{2}]-dt^{2},  \label{flrw}
\end{equation}%
where the scale factor $\mathring{a}(t)$ (we use also the value $\mathring{H}%
:=\mathring{a}/\mathring{a},$ for $\mathring{a}^{\diamond }:=d\mathring{a}%
/dt)$ with signature $(+,+,+,-)$, and a parametrization of coordinates in
the form $u^{\alpha ^{\prime }}=(x^{1^{\prime }}=r,x^{2^{\prime }}=\theta
,y^{3^{\prime }}=\varphi ,y^{4^{\prime }}=t),$ or as Cartesian coordinates $%
(x^{1^{\prime }}=x,x^{2^{\prime }}=y,y^{3^{\prime }}=z,y^{4^{\prime }}=t).$
For such cosmological metrics, the main issues of the Einstein and modified
Universes are encoded into energy-momentum tensor $\mathring{T}_{\alpha
\beta }=\mathring{p}\mathring{g}_{\alpha \beta }+(\mathring{\rho}+\mathring{p%
})\mathring{v}_{\alpha }\mathring{v}_{\beta }$ (we omit primes or other
distinctions in the coordinate indices if there is no ambiguity) arising
from a matter Lagrangian $\ ^{m}\mathcal{\mathring{L}}$ for $\mathring{T}_{\
\beta }^{\alpha }=diag[0,0,0,-\mathring{\rho}]$ with
\begin{equation}
\mathring{T}(t)=\mathring{T}_{\ \alpha }^{\alpha }=-\mathring{\rho},%
\mathring{P}(t)=\mathring{R}_{\alpha \beta }\mathring{T}^{\alpha \beta }=%
\mathring{R}_{44}\mathring{T}^{44}=-3\mathring{\rho}(\mathring{H}^{2}+%
\mathring{H}^{\diamond }).  \label{auxaa}
\end{equation}

There will constructed nonhomogeneous and locally anisotropic cosmological
solutions of type \ (\ref{dm1}) with
\begin{eqnarray}
g_{i} &=&g_{i}{(x^{k})}=\eta _{i}(x^{k},y^{4})\mathring{g}%
_{i}(x^{k},y^{4})=e^{\psi {(x^{k})}},  \label{polarf} \\
g_{a} &=&\omega ^{2}(x^{k},y^{4})h_{a}(x^{k},y^{4})=\omega
^{2}(x^{k},y^{4})\eta _{a}(x^{k},y^{4})\mathring{g}_{a}(x^{k},y^{4}),\ \
N_{i}^{3}=n_{i}(x^{k}),N_{i}^{4}=w_{i}(x^{k},y^{4}).  \notag
\end{eqnarray}%
In Eqs.~(\ref{polarf}), there is no summation on repeated indices, $\eta
_{\alpha }=(\eta _{i},\eta _{a})$ are polarization functions, the
N-connection coefficients are determined by $n_{i}$ and $w_{i},$ the
vertical conformal factor $\omega $ may depend on all spacetime coordinates
and $\mathring{g}_{\alpha }=(\mathring{g}_{i},\mathring{g}_{a})$ define the
\textquotedblleft prime\textquotedblright\ diagonal metric if $\eta _{\alpha
}=1$ and $N_{i}^{a}=0.$ The \textquotedblleft target\textquotedblright\
off-diagonal metrics are with Killing symmetry on $\partial /\partial y^{3}$
when the coefficients (\ref{polarf}) do not depend on $y^{3}.$ We can
consider nonholonomic deformations with non-Killing symmetries when, for
instance, $\omega (x^{k},y^{4})\rightarrow \omega (x^{k},y^{3},y^{4})$,
which results in a more cumbersome calculus and geometric techniques. For
simplicity, we do not study such generalizations in this work (see examples
in \cite{voffds}).

The quadratic line element is parameterized
\begin{equation}
ds^{2}=a^{2}(x^{k},t)[\eta _{1}(x^{k},t)(dx^{1})^{2}+\eta
_{2}(x^{k},t)(dx^{2})^{2}]+a^{2}(x^{k},t)\widehat{h}_{3}(x^{k},t)(\widehat{%
\mathbf{e}}^{3})^{2}+\omega ^{2}(x^{k},t)h_{4}(x^{k},t)(\widehat{\mathbf{e}}%
^{4})^{2},  \label{flrwod}
\end{equation}%
when $a^{2}(x^{k},t)\eta _{i}(x^{k},t)=e^{\psi {(x^{k})}},$ for $i=1,2;$ $%
a^{2}\ \widehat{h}_{3}=\omega ^{2}(x^{k},t)h_{3}(x^{k},t),$ and $\ \widehat{%
\mathbf{e}}^{3}=dy^{3}+n_{i}(x^{k})dx^{i},\widehat{\mathbf{e}}%
^{4}=dy^{4}+w_{i}(x^{k},t)dx^{i}.$ Functions $\eta _{i},\eta _{a},a,\psi
,\omega ,n_{i},w_{i}$ will be found such that, via nonholonomic transforms (%
\ref{polarf}), when $\mathring{g}_{\alpha ^{\prime }\beta ^{\prime }}(t)$ (%
\ref{flrw}) $\rightarrow $ $\widehat{\mathbf{g}}_{\alpha \beta }(x^{k},t)$ (%
\ref{flrwod}), off-diagonal nonhomogeneous cosmological solutions are
generated in a model of MGT (\ref{mgts}). We can consider subclasses of
off-diagonal cosmological solutions with deformed symmetries when nontrivial
limits $\widehat{\mathbf{g}}_{\alpha \beta }(x^{k},t)\rightarrow $ $\widehat{%
\mathbf{g}}_{\alpha \beta }(t)$ can be found and define viable cosmological
models.

Applying an N-adapted variational procedure with respect to nonholonomic
bases (\ref{nadif}) for the action $\mathcal{S}=\ ^{g}\widehat{\mathbf{S}}+\
^{m}\widehat{\mathbf{S}},$ which is similar to that in \cite{odgom} but for $%
\nabla \rightarrow $ $\widehat{\mathbf{D}}$ and matter source $\widehat{%
\mathbf{T}}_{\alpha \beta }$ (\ref{dsourc}), we obtain the field equations
for the corresponding modified gravity theory
\begin{eqnarray}
&&\ \widehat{\mathbf{R}}_{\alpha \beta }\ \ ^{1}\widehat{\mathbf{f}}-\frac{1%
}{2}\ \widehat{\mathbf{g}}_{\alpha \beta }\widehat{\mathbf{f}}+(\widehat{%
\mathbf{g}}_{\alpha \beta }\widehat{\mathbf{D}}^{\mu }\widehat{\mathbf{D}}%
_{\mu }-\widehat{\mathbf{D}}_{\alpha }\widehat{\mathbf{D}}_{\beta })\ ^{1}%
\widehat{\mathbf{f}}+(\widehat{\mathbf{T}}_{\alpha \beta }+\mathbf{\Theta }%
_{\alpha \beta })\ ^{2}\widehat{\mathbf{f}}+  \label{mgtfe} \\
&&\mathbf{\Xi }_{\alpha \beta }\ ^{3}\widehat{\mathbf{f}}+\frac{1}{2}(%
\widehat{\mathbf{D}}^{\mu }\widehat{\mathbf{D}}_{\mu }\widehat{\mathbf{T}}%
_{\alpha \beta }\ ^{3}\widehat{\mathbf{f}}+\widehat{\mathbf{g}}_{\alpha
\beta }\widehat{\mathbf{D}}_{\mu }\widehat{\mathbf{D}}_{\nu }\widehat{%
\mathbf{T}}^{\mu \nu }\ ^{3}\widehat{\mathbf{f}})-\widehat{\mathbf{D}}_{\nu }%
\widehat{\mathbf{D}}_{(\alpha }\widehat{\mathbf{T}}_{\beta )}^{\ \nu }\ ^{3}%
\widehat{\mathbf{f}}=\kappa ^{2}\ \widehat{\mathbf{T}}_{\alpha \beta },
\notag
\end{eqnarray}
for $\mathbf{\Theta }_{\alpha \beta }=p\ \widehat{\mathbf{g}}_{\alpha \beta
}-2\widehat{\mathbf{T}}_{\alpha \beta },\ \mathbf{\Xi }_{\alpha \beta }=2\
\widehat{\mathbf{E}}_{\ (\alpha }^{\nu }\widehat{\mathbf{T}}_{\beta )\nu
}-p\ \widehat{\mathbf{E}}_{\alpha \beta }-\frac{1}{2}\widehat{\mathbf{R}}%
\widehat{\mathbf{T}}_{\alpha \beta }$, with respective d-tensors defined by
Eqs. (\ref{driccic}), where $\ ^{1}\widehat{\mathbf{f}}:=\partial \widehat{%
\mathbf{f}}/\partial \widehat{\mathbf{R}},$ $\ \ \ ^{2}\widehat{\mathbf{f}}%
:=\partial \widehat{\mathbf{f}}/\partial \widehat{\mathbf{T}}$ and $\ \ ^{3}%
\widehat{\mathbf{f}}:=\partial \widehat{\mathbf{f}}/\partial \widehat{%
\mathbf{P}},$ when $\widehat{\mathbf{P}}=\widehat{\mathbf{R}}_{\alpha \beta }%
\widehat{\mathbf{T}}^{\alpha \beta }$ and $(\alpha \beta ) $ denotes
symmetrization of the indices.

In general, the divergence with $\widehat{\mathbf{D}}$ and/or $\nabla $ of
Eqs.~(\ref{mgtfe}) is not zero. Also Eqs.~(\ref{nheeq}) have a similar
property. In the last case, we can obtain the continuity equations as in GR
and then deform them by using the distortions (\ref{distr}), which for the
canonical d-connections are completely determined by the metric structure.
There are certain types of conservation laws for matter fields with
additional nonholonomic constraints. Remarkably, such sophisticate
nonholonomic and nonlinear systems can be solved in very general
off-diagonal forms, by applying the anholonomic frame deformation method. In
order to compare these results and to find possible applications in modern
cosmology, we will consider a particular equation of state (EoS) $p=\varpi
\rho $ with $\varpi =const,$ and study the cosmology of off-diagonal
distortions of certain FLRW models considered in the framework of GR and its
modifications. In both cases, by exploring some particular classes of
solutions, the dynamics of the matter sector of generalized $f(R,T,R_{\mu
\nu }T^{\mu \nu })$\ gravity (with respect to N-adapted frames) may lead to
similar cosmological scenarios as GR, but with nonholonomic constraints and
deformations.

\section{ The AFDM\ and exact solutions in MGTs}

\label{safdm} A surprising property of Eqs.~(\ref{nheeq}) and (\ref{mgtfe})
is that they can be integrated in very general form with generic
off-diagonal metrics when their coefficients depend on all spacetime
coordinates via various classes of generating and integration functions and
constants. In particular, we can consider such generating and integration
functions when $\widehat{\mathbf{g}}_{\alpha \beta }(x^{k},t)$ (\ref{flrwod}%
) result in off-diagonal metrics of type $\widehat{\mathbf{g}}_{\alpha \beta
}(t)$ depending on the parameters and possible (non-) commutative Lie
algebra or algebroid symmetries.

\subsection{Off-diagonal FLRW like cosmological models}

We shall study cosmological models with sources of type (\ref{dsourc}) when
the four-velocity $\widehat{\mathbf{v}}_{\alpha }$ is re-parameterized in a
way that for some frame transforms as
\begin{eqnarray}
\widehat{\mathcal{Y}}_{\alpha \beta }:=\kappa ^{2}\ (\widehat{\mathbf{T}}%
_{\alpha \beta }-\frac{1}{2}\mathbf{g}_{\alpha \beta }\widehat{\mathbf{T}})
&\rightarrow &diag[\Upsilon _{1}=\Upsilon _{2},\Upsilon _{2}=\ ^{h}\Upsilon
(x^{i}),\Upsilon _{3}=\Upsilon _{4},\Upsilon _{4}=\ ^{v}\Upsilon (x^{i},t)]
\label{dsours1} \\
&\rightarrow &\widehat{\Lambda }\ \mathbf{g}_{\alpha \beta }\
\mbox{
(redefining the generating functions and sources)},  \label{dsours2}
\end{eqnarray}%
for effective $h$- and $v$-polarized sources, respectively, $\ ^{h}\Upsilon
(x^{i})$ and $\Upsilon _{4}=\ ^{v}\Upsilon (x^{i},t),$ or an effective
cosmological constant $\widehat{\Lambda }.$ For simplicity, we can consider
effective matter sources and ``prime'' metrics with Killing symmetry on $%
\partial /\partial _{3},$ i.e. when the effective matter sources and
d-metrics do not depend on $y^{3}.$\footnote{%
The method can be extended to account for $y^{3}$ dependence and non-Killing
configurations (see \cite{voffds}). In this paper the local coordinates and
ans\"{a}tze for d-metrics are parameterized in different forms than in
previous works, what is more convenient for the study of cosmological models.%
} In brief, the partial derivatives $\partial _{\alpha }=\partial /\partial
u^{\alpha }$ will be written as $s^{\bullet }=\partial s/\partial
x^{1},s^{\prime }=\partial s/\partial x^{2},s^{\ast }=\partial s/\partial
y^{3},s^{\diamond }=\partial s/\partial y^{4}.$

The nontrivial components of the nonholonomic Einstein equations (\ref%
{driccic}), with source (\ref{dsours1}) parameterized with respect to (co)
bases (\ref{nadif}), for a d-metric ans\"{a}tze (\ref{dm1}) with
coefficients (\ref{polarf}), are (see Refs. \cite{odgom})
\begin{eqnarray}
\psi ^{\bullet \bullet }+\psi ^{\prime \prime } &=&2~^{h}\Upsilon
\label{eq1m} \\
\phi ^{\diamond }h_{3}^{\diamond } &=&2h_{3}h_{4}~^{v}\Upsilon  \label{eq2m}
\\
n_{i}^{\diamond \diamond }+\gamma n_{i}^{\diamond } &=&0,  \label{eq3m} \\
\beta w_{i}-\alpha _{i} &=&0,  \label{eq4m} \\
\partial _{i}\omega -(\partial _{i}\phi /\phi ^{\diamond })\omega ^{\diamond
} &=&0,  \label{confeq}
\end{eqnarray}%
for
\begin{equation}
\alpha _{i}=h_{3}^{\diamond }\partial _{i}\phi ,\beta =h_{3}^{\diamond }\
\phi ^{\diamond },\gamma =\left( \ln |h_{3}|^{3/2}/|h_{4}|\right) ^{\diamond
},  \label{abc}
\end{equation}%
where
\begin{equation}
{\phi =\ln |h_{3}^{\diamond }/\sqrt{|h_{3}h_{4}|}|,\mbox{ and/ or }}\Phi
:=e^{{\phi }},  \label{genf}
\end{equation}%
is considered as a generating function. In these formulas, we consider $%
h_{a}^{\diamond }\neq 0,$ $\ ^{h}\Upsilon ,\ ^{v}\Upsilon \neq 0.$ Formula (%
\ref{confeq}) is a nontrivial solution of (\ref{eq4m}) with coefficients (%
\ref{abc}), when
\begin{equation}
w_{i}=\partial _{i}\phi /\phi ^{\diamond }  \label{aux4}
\end{equation}
and  $\mathbf{e}_{i}\omega =\partial _{i}\omega -n_{i}\ \omega ^{\ast
}-w_{i}\omega ^{\diamond }=0$.

The d-torsion (\ref{dtors}) vanishes if the (Levi-Civita, LC) conditions $%
\widehat{L}_{aj}^{c}=e_{a}(N_{j}^{c}),\widehat{C}_{jb}^{i}=0,\Omega _{\
ji}^{a}=0,$ are satisfied:
\begin{eqnarray}
w_{i}^{\diamond } &=&(\partial _{i}-w_{i}\partial _{4})\ln \sqrt{|h_{4}|}%
,(\partial _{i}-w_{i}\partial _{4})\ln \sqrt{|h_{3}|}=0,  \label{lccondb} \\
\partial _{k}w_{i} &=&\partial _{i}w_{k},n_{i}^{\diamond }=0,\partial
_{i}n_{k}=\partial _{k}n_{i}.  \notag
\end{eqnarray}

The decoupling property of the above system of equations follows from the
facts that: 1) integrating the 2-d Laplace equation (\ref{eq1m}) one finds
solutions for the $h$-coefficients of the d-metric, and 2) the solutions for
the coefficients of the d-metric can be found from (\ref{eq2m}) and (\ref%
{genf}). \ 3) Then the N-connection coefficients $w_{i}$ and $n_{i}$ can be
found from (\ref{eq3m}) and (\ref{eq4m}), respectively.

\subsubsection{Cosmological solutions with nonholonomically induced torsion}

The equations (\ref{eq1m}) and (\ref{eq4m}) can be solved, respectively, for
any source $~^{h}\Upsilon (x^{k})$ and generating function ${\phi (x}^{k},t{%
).}$ The system (\ref{eq2m}) and (\ref{genf}) can be written as $%
h_{3}h_{4}=\phi ^{\diamond }h_{3}^{\diamond }/2~^{v}\Upsilon $ and $%
|h_{3}h_{4}|=({h_{3}^{\diamond })}^{2}e^{-2\phi },$ for any nontrivial
source $~^{v}\Upsilon (x^{i},t)$ in (\ref{eq2m}). Introducing the first
equation into the second, one finds $|h_{3}^{\diamond }|=\frac{(e^{2\phi
})^{\diamond }}{4|~^{v}\Upsilon |}=\frac{\Phi ^{\diamond }\Phi \ }{%
2|~^{v}\Upsilon |},$ i.e. $h_{3}=\ ^{0}h_{3}(x^{k})+\frac{\epsilon
_{3}\epsilon _{4}}{4}\int dt\frac{(\Phi ^{2})^{\diamond }}{~^{v}\Upsilon }$,
where $\ ^{0}h_{3}(x^{k})$ and $\epsilon _{3},\epsilon _{4}=\pm 1.$ Using
again the 1st eq., we get $\ h_{4}=\frac{\phi ^{\diamond }(\ln \sqrt{|h_{3}|}%
)^{\diamond }}{2~^{v}\Upsilon }=\frac{1}{2~^{v}\Upsilon }\frac{\Phi
^{\diamond }}{\Phi }\frac{h_{3}^{\diamond }}{h_{3}}.$ We can simplify such
formulas for $h_{3}$ and $h_{4}$ if we redefine the generating function, $%
\Phi \rightarrow \widehat{\Phi },$ where $(\Phi ^{2})^{\diamond
}/|~^{v}\Upsilon |=(\widehat{\Phi }^{2})^{\diamond }/\Lambda ,$ i.e.%
\begin{equation}
\Phi ^{2}=\Lambda ^{-1}\left[ \widehat{\Phi }^{2}|~^{v}\Upsilon |+\int dt\
\widehat{\Phi }^{2}|~^{v}\Upsilon |^{\diamond }\right] ,  \label{aux2}
\end{equation}%
for an effective cosmological constant $\Lambda $ which may take positive or
negative values. We can integrate on $t,$ include the integration function $%
\ ^{0}h_{3}(x^{k})$ in $\widehat{\Phi }$ and write
\begin{equation}
h_{3}[\widehat{\Phi }]=\widehat{\Phi }^{2}/4\Lambda .  \label{h3}
\end{equation}%
Introducing this formula and (\ref{aux2}) and that for $h_{4},$ we compute
\begin{equation}
h_{4}[\widehat{\Phi }]=\frac{(\ln |\Phi |)^{\diamond }}{4|~^{v}\Upsilon |}=%
\frac{(\widehat{\Phi }^{2})^{\diamond }}{8}\left[ \widehat{\Phi }%
^{2}|~^{v}\Upsilon |+\int dt\ \widehat{\Phi }^{2}|~^{v}\Upsilon |^{\diamond }%
\right] ^{-1}.  \label{h4}
\end{equation}

As next step, we need solve Eq.~(\ref{eq3m}) by integrating on $t$ twice. We
obtain
\begin{equation}
n_{k}=\ _{1}n_{k}+\ _{2}n_{k}\int dt\ h_{4}/(\sqrt{|h_{3}|})^{3},
\label{n1b}
\end{equation}%
where $\ _{1}n_{k}(x^{i}),\ _{2}n_{k}(x^{i})$ are integration functions and $%
h_{a}[\widehat{\Phi }]$ are given by formulas (\ref{h3}) and (\ref{h4}). If
we fix $\ _{2}n_{k}=0,$ we shall be able to find $n_{k}=\ _{1}n_{k}(x^{i})$
which have zero torsion limits (see examples in subsection \ref{sslc}).

The solutions of (\ref{eq4m}) are given by (\ref{aux4}), which for different
types of generating functions are parameterized
\begin{equation}
w_{i}=\frac{\partial _{i}\Phi }{\Phi ^{\diamond }}=\frac{\partial _{i}(\Phi
^{2})}{(\Phi ^{2})^{\diamond }},  \label{w1b}
\end{equation}%
where the integral functional $\Phi \lbrack \widehat{\Phi },~^{v}\Upsilon ]$
is given by (\ref{aux2}).

We can introduce certain polarization functions $\eta _{\alpha }$ in order
to write the d-metric of such solutions in the form (\ref{flrwod}). Let us
fix $\omega ^{2}=|h_{4}|^{-1}$ to satisfy the condition (\ref{confeq}),
which for a generating function $\Phi \lbrack \phi ]$ is equivalent to  $%
\Phi ^{\diamond }\partial _{i}h_{4}-\partial _{i}\Phi \ h_{4}^{\diamond }=0$%
.  These first order PDE equations impose certain conditions on the class of
generating function $\Phi $ and source $~^{v}\Upsilon .$ We can choose such
a system of coordinates where $~^{v}\Upsilon =\frac{1}{4}(e^{-\phi
})^{\diamond }$ and $h_{4}=\Phi ,$ i.e. this coefficient of the d-metric is
considered as a generating function and the last conditions are solved.

A modification of the scale factor $\mathring{a}(t)\rightarrow a(x^{k},t),$
for the FLRW metric (\ref{flrw}) (with for $\mathring{g}_{1}=\mathring{g}%
_{2}=\mathring{g}_{3}=\mathring{a}^{2},\mathring{g}_{4}=-1$, has to be
chosen in order to explain observational cosmological data. For any
prescribed functions $a(x^{k},t)$ and $\omega ^{2}=|h_{4}|^{-1}$ and
solutions \ $e^{\psi {(x^{k})}},$ (see (\ref{eq1m})) and $h_{a}[\widehat{%
\Phi }],n_{k}(x^{i}),w_{i}[\widehat{\Phi }]$ (given respectively by formulas
(\ref{h3})-(\ref{w1b})), we can compute the polarization functions $\eta
_{i}=a^{-2}e^{\psi },\eta _{3}=\mathring{a}^{-2}h_{3},\eta _{4}=1$ and
function $\widehat{h}_{3}=h_{3}/a^{2}|h_{4}|.$ Such coefficients (see the
data (\ref{polarf})), define off-diagonal metrics of type (\ref{flrwod}),
\begin{equation}
ds^{2} =a^{2}(x^{k},t)[\eta _{1}(x^{k},t)(dx^{1})^{2}+\eta
_{2}(x^{k},t)(dx^{2})^{2}]+ a^{2}(x^{k},t)\widehat{h}%
_{3}(x^{k},t)[dy^{3}+n_{i}(x^{k})dx^{i}]^{2}-[dt+\frac{\partial _{i}\Phi
\lbrack \widehat{\Phi },~^{v}\Upsilon ]}{\Phi ^{\diamond }[\widehat{\Phi }%
,~^{v}\Upsilon ]}dx^{i}]^{2}.  \label{odfrlwtors}
\end{equation}%
Choosing any generating functions $a^{2}(x^{k},t),\psi (x^{i})$ and $\Phi
\lbrack \widehat{\Phi },~^{v}\Upsilon ]$ and integration functions $%
n_{i}(x^{k}),$ we generate a nonhomogeneous cosmological model with
nonholonomically induced torsion (\ref{dtors}). More general torsions can be
induced if $n_{i}(x^{k},t)$ is taken with two types of integration functions
$\ _{1}n_{i}(x^{k})$ and $\ _{2}n_{i}(x^{k})$ (see Eqs.~(\ref{n1b})). Having
constructed this solution, we can now consider certain subclasses of
generating and integration functions where $a(x^{k},t)\rightarrow a(t)\neq
\mathring{a}(t),w_{i}\rightarrow w_{i}(t),n_{i}\rightarrow const$, etc. In
this way generic off-diagonal cosmological metrics are generated (because
there are nontrivial anholonomy coefficients $W_{ia}^{b}$.

\subsubsection{Levi-Civita off-diagonal cosmological configurations}

\label{sslc}

The LC-conditions (\ref{lccondb}) are given by a set of nonholonomic
constraints which cannot be solved in explicit form for arbitrary data $%
(\Phi ,\Upsilon )$ and arbitrary integration functions $\ _{1}n_{k}$ and $\
_{2}n_{k}.$ However, some subclasses of off-diagonal solutions can still be
constructed where via frame and coordinate transforms we can chose $\
_{2}n_{k}=0$ and $\ _{1}n_{k}=\partial _{k}n$ with a function $n=n(x^{k}).$
It should be noted that $(\partial _{i}-w_{i}\partial _{4})\Phi \equiv 0$
for any $\Phi (x^{k},y^{4})$ if $w_{i}$ is defined by (\ref{w1b}).
Introducing a new functional $B(\Phi ),$ we find that $(\partial
_{i}-w_{i}\partial _{4})B=\frac{\partial B}{\partial \Phi }(\partial
_{i}-w_{i}\partial _{4})\Phi =0.$ Using Eq.~(\ref{h3}) for functionals of
type $h_{3}=B(|\tilde{\Phi}(\Phi )|),$ we solve Eqs.~$(\partial
_{i}-w_{i}\partial _{4})h_{3}=0,$ what is equivalent to the second system of
equations in (\ref{lccondb}), because $(\partial _{i}-w_{i}\partial _{4})\ln
\sqrt{|h_{3}|}\sim (\partial _{i}-w_{i}\partial _{4})h_{3}.$

We can use a subclass of generating functions $\Phi =\check{\Phi}$ for which
$\ (\partial _{i}\check{\Phi})^{\diamond }=\partial _{i}\check{\Phi}%
^{\diamond }$ and get for the left part of the second equation in (\ref%
{lccondb}), $(\partial _{i}-w_{i}\partial _{4})\ln \sqrt{|h_{3}|}=0.$ The
first system of equations in (\ref{lccondb}) can be solved in explicit form
if $w_{i}$ are determined by formulas (\ref{w1b}), and $h_{3}[\tilde{\Phi}]$
and $h_{4}[\tilde{\Phi},\tilde{\Phi}^{\diamond }]$ are chosen respectively
for any $\Upsilon \rightarrow \Lambda .$ We can consider $\tilde{\Phi}=%
\tilde{\Phi}(\ln \sqrt{|h_{4}|})$ for a functional dependence $h_{4}[\tilde{%
\Phi}[\check{\Phi}]].$ This allows us to obtain $w_{i}=\partial _{i}|\tilde{%
\Phi}|/|\tilde{\Phi}|^{\diamond }=\partial _{i}|\ln \sqrt{|h_{4}|}|/|\ln
\sqrt{|h_{4}|}|^{\diamond }.$ Taking the derivative $\partial _{4}$ on both
sides of these equations, we get  $w_{i}^{\diamond }=\frac{(\partial
_{i}|\ln \sqrt{|h_{4}|}|)^{\diamond }}{|\ln \sqrt{|h_{4}|}|^{\diamond }}%
-w_{i}\frac{|\ln \sqrt{|h_{4}|}|^{^{\diamond }}}{|\ln \sqrt{|h_{4}|}%
|^{^{\diamond }}}$. If the mentioned conditions are satisfied, we can
construct in explicit form generic off-diagonal configurations with $%
w_{i}^{^{\diamond }}=(\partial _{i}-w_{i}\partial _{4})\ln \sqrt{|h_{4}|},$
which is necessary for the zero torsion conditions. \ We need also to solve
for the conditions $\partial _{k}w_{i}=\partial _{i}w_{k}$ from the second
line in (\ref{lccondb}). We find in explicit form the solutions for such
coefficients,
\begin{equation}
\check{w}_{i}=\partial _{i}\check{\Phi}/\check{\Phi}^{^{\diamond }}=\partial
_{i}\widetilde{A},  \label{w1c}
\end{equation}%
with a nontrivial function $\widetilde{A}(x^{k},y^{4})$ depending
functionally on the generating function $\check{\Phi}.$

Finally, we conclude that we generate LC-configurations for a class of
off-diagonal cosmological metric type (\ref{dm1}) for $\Upsilon =\breve{%
\Upsilon}=\Lambda ,$ $\Phi =\check{\Phi}=\tilde{\Phi}$ and $\ _{2}n_{k}=0$
in (\ref{n1b}) which are parameterized by quadratic elements
\begin{equation}
ds^{2}=e^{\psi (x^{k})}[(dx^{1})^{2}+(dx^{2})^{2}]+\frac{\check{\Phi}^{2}}{%
4|\Lambda |}[dy^{3}+(\partial _{k}n(x^{i}))dx^{k}]^{2}-\frac{(\check{\Phi}%
^{^{\diamond }})^{2}}{|\Lambda |\check{\Phi}^{2}}[dt+(\partial _{i}%
\widetilde{A}[\check{\Phi}])dx^{i}]^{2}.  \label{qelgen}
\end{equation}%
We can re-write such solutions in the form (\ref{odfrlwtors}). This provides
us a general procedure of off-diagonal deformations with $\mathring{a}%
(t)\rightarrow a(x^{k},t)$ (see the FLRW metric (\ref{flrw})), resulting in
nonhomogeneous cosmological metrics in GR. Prescribing a function $%
a(x^{k},t),$ a generating function $\check{\Phi}(x^{k},t)$ and a solution $%
e^{\psi {(x^{k})}}$ (see (\ref{eq1m})), we respectively compute the
v-conformal factor and the polarization functions for  $\widehat{h}%
_{3}=h_{3}/a^{2}|h_{4}|=\check{\Phi}^{4}/4a^{2}(\check{\Phi}^{^{\diamond
}})^{2},\ \omega ^{2}=|h_{4}|^{-1}=|\Lambda |\check{\Phi}^{2}/(\check{\Phi}%
^{^{\diamond }})^{2},\ \eta _{i}=a^{-2}e^{\psi },\ \eta _{3}=\mathring{a}%
^{-2}h_{3}=\check{\Phi}^{2}/4|\Lambda |\mathring{a}^{2},\ \eta _{4}=1$.
Such coefficients (see data (\ref{polarf})) transform the off-diagonal
cosmological solutions (\ref{qelgen}) into metrics of type (\ref{flrwod}),
\begin{equation}
ds^{2}=a^{2}(x^{k},t)\{[\eta _{1}(x^{k},t)(dx^{1})^{2}+\eta
_{2}(x^{k},t)(dx^{2})^{2}]+\widehat{h}_{3}(x^{k},t)[dy^{3}+(\partial
_{k}n(x^{i}))dx^{k}]^{2}\}-[dt+(\partial _{i}\widetilde{A}[\check{\Phi}%
])dx^{i}]^{2}.  \label{qelgenofd}
\end{equation}%
The dependence on the source $\Lambda $ is contained in explicit form, for
instance, in the polarization $\eta _{3}$. This class of effective Einstein
off-diagonal metrics $\mathbf{g}_{\alpha \beta }(x^{k},t)$ define new
nonhomogeneous cosmological solutions in GR as off-diagonal deformations of
the FLRW cosmology. For certain well-defined conditions, one can find limits
$\mathbf{g}_{\alpha \beta }\rightarrow \mathbf{g}_{\alpha \beta }(t,a(t),%
\widehat{h}_{3}(t),\check{\Phi}(t),\eta _{i}(t)).$ This provides explicit
geometric models of nonlinear off-diagonal anisotropic cosmological
evolution which, with respect to N-adapted frames, describe $a(t)$ with
modified re-scaling factors.

\subsection{Effective FLRW cosmology for $f$-modified gravity}

\label{sseflrw}

The anholonomic frame deformation method outlined in previous subsections
can be applied for the generation of off-diagonal cosmological solutions of
field equations of modified gravities, see (\ref{mgtfe}). Redefining the
generating functions via the transforms (\ref{aux2}) and $\Phi \rightarrow
\check{\Phi}\rightarrow \tilde{\Phi}$, we can generate off-diagonal
cosmological configurations with $\ \widehat{\mathbf{R}}=4\Lambda ,$ see (%
\ref{dsours1}) and (\ref{dsours2}). Such parameterizations of geometric data
and sources are possible for certain general conditions via transforms of
N-adapted frames when the action functional functionally depends on $\Lambda
$ and on the effective sources, $\widehat{\mathbf{f}}[\widehat{\mathbf{R}}%
(\Lambda ),\widehat{\mathbf{T}}(\Lambda ),\widehat{\mathbf{P}}],$ with $%
\widehat{\mathbf{P}}(t)=\widehat{\mathbf{R}}_{\alpha \beta }\widehat{\mathbf{%
T}}^{\alpha \beta }=-3\mathring{\rho}(H^{2}+H^{\diamond })\ $ and $%
H=a^{\diamond }/a~\ $with scaling factor $a(t)$ taken for some limits of a
solution (\ref{odfrlwtors}), or (\ref{qelgenofd}).

We assume that the density of matter $\rho =\mathring{\rho}$ in $\widehat{%
\mathbf{T}}_{\alpha \beta }$ (\ref{dsourc}) is the same as for a standard
FLRW metric (\ref{flrw}) and does not change under off-diagonal deformations
with respect to N-adapted frames. For such configurations, \ $\mathbf{\Theta
}_{\ \beta }^{\alpha }=(p\ -2\Lambda )\delta _{\ \beta }^{\alpha }$ and $\
\mathbf{\Xi }_{\ \beta }^{\alpha }=(2\Lambda ^{2}-p\ \Lambda -\frac{1}{2}%
4\Lambda ^{2})\delta _{\ \beta }^{\alpha }=-p\Lambda \delta _{\ \beta
}^{\alpha },$ where terms with $\Lambda ^{2}$ compensate each other in 4-d.
We can write $\widehat{\mathbf{D}}_{\mu }\widehat{\mathbf{T}}_{\alpha \beta
}=0,$ $\widehat{\mathbf{D}}_{\mu }\ ^{1}\widehat{\mathbf{f}}\sim \partial
^{2}\widehat{\mathbf{f}}/\partial \widehat{\mathbf{R}}_{\ldots }^{2}$ $%
\mathbf{e}_{\mu }\Lambda \sim 0,$ and (similarly) $\widehat{\mathbf{D}}_{\mu
}\ ^{2}\widehat{\mathbf{f}}\sim 0,$ $\widehat{\mathbf{D}}_{\mu }\ ^{3}%
\widehat{\mathbf{f}}\sim 0,$ for \ $\widehat{\mathbf{R}}_{\alpha \beta }\sim
$ $\widehat{\mathbf{T}}_{\alpha \beta }\sim \Lambda \delta _{\alpha \beta },$
$\Lambda =const,$ with respect to corresponding classes of N-adapted frames.
Eqs.~(\ref{mgtfe}) transform into a system of nonholonomic nonlinear PDEs of
type (\ref{eq1m})-(\ref{eq4m}), $\ \widehat{\mathbf{R}}_{\ \ \beta }^{\alpha
}=\ \widehat{\mathbf{\Upsilon }}\delta _{\ \beta }^{\alpha }$, with
effective diagonalized source
\begin{equation}
\ \widehat{\mathbf{\Upsilon }}=\frac{\Lambda }{\ ^{1}\widehat{\mathbf{f}}}+%
\frac{\widehat{\mathbf{f}}}{2\ ^{1}\widehat{\mathbf{f}}}+(2\Lambda -\kappa
^{-2}\Lambda -p)\frac{\ ^{2}\widehat{\mathbf{f}}}{\ ^{1}\widehat{\mathbf{f}}}%
+p\Lambda \frac{\ ^{3}\widehat{\mathbf{f}}}{\ ^{1}\widehat{\mathbf{f}}},
\label{effsource}
\end{equation}%
which can be parameterized with dependencies on $(x^{i},t),$ or on $t.$
These equations can be solved for very general off-diagonal forms, depending
on generating and integration functions, following the procedure outlined in
previous subsections. Redefining the generation function as in (\ref{aux2}),
when an effective cosmological constant $\check{\Lambda}$ is generated from $%
\widehat{\mathbf{\Upsilon }}(x^{i},t)$, one has $\check{\Phi}^{2}=\check{%
\Lambda}^{-1}\left[ \widehat{\Phi }^{2}|\ \widehat{\mathbf{\Upsilon }}|+\int
dt\ \widehat{\Phi }^{2}|\ \widehat{\mathbf{\Upsilon }}|^{\diamond }\right]$.
Such a generating function defines off-diagonal cosmological solutions of
type (\ref{odfrlwtors}), or (\ref{qelgen}), as solutions of field equations
for an effective (nonholonomic) Einstein space $\mathbf{\check{R}}_{\ \beta
}^{\alpha }=\check{\Lambda}\delta _{\ \beta }^{\alpha }$. In this way, a
geometric method is provided when the (effective or modified) matter sources
transform as $\ \widehat{\mathbf{\Upsilon }}$ (\ref{dsours1}) $\rightarrow $
$\check{\Lambda}$ (\ref{dsours2}) and the gravitational field equations in
modified gravity can be effectively expressed as nonholonomic Einstein
spaces when the d-metric coefficients encode the contributions of $\widehat{%
\mathbf{f}},\ ^{1}\widehat{\mathbf{f}},\ ^{2}\widehat{\mathbf{f}}$ and $\
^{3}\widehat{\mathbf{f}}$ and of the matter sources.

We can consider inverse transforms with $\ \check{\Lambda}\rightarrow
\widehat{\mathbf{\Upsilon }}$ and state that for certain well-defined
conditions (\ref{w1c}) we can mimic both $f$-functional contributions and/or
massive gravitational theories \cite{voffdmgt}. Here we emphasize that
off-diagonal configurations (of vacuum and non-vacuum types) are possible
even if the effective sources from modified gravity are constrained to be
zero.

\section{Off-diagonal modeling of cosmological modified gravity theories}

\label{s4} This section has three goals. The first is to provide a
reconstruction procedure for off-diagonal effective Einstein and modified
gravity cosmological scenarios. The second is to apply these methods in
practice and provide explicit examples related to $f(R)$ gravity and
cosmology. The third goal is to analyze how matter stability problems for $%
f(R)$-theories can be solved by nonholonomic frame transforms and
deformations and imposing non-integrable constraints.

\subsection{Reconstructing nonholonomic $f$-models}

Let us construct an effective Einstein space which models a quite general
modified gravity theory with $f(R,T,R_{\alpha \beta }T^{\alpha \beta
})=R+F(R_{\alpha \beta }T^{\alpha \beta })+G(T).$ This theory admits a
reconstruction procedure which is similar to that elaborated in \cite{odgom}. Following the
anholonomic frame deformation method with an auxiliary canonical
d-connection $\widehat{\mathbf{D}},$ the modified gravity (\ref{mgtfe}) is
formulated for
\begin{equation}
\widehat{\mathbf{f}}(\widehat{\mathbf{R}},\widehat{\mathbf{T}},\widehat{%
\mathbf{R}}_{\alpha \beta }\widehat{\mathbf{T}}^{\alpha \beta })=\widehat{%
\mathbf{R}}+\widehat{\mathbf{F}}(\widehat{\mathbf{P}})+\widehat{\mathbf{G}}(%
\widehat{\mathbf{T}}).  \label{aux10}
\end{equation}%
We can self-consistently embed this model into a nonholonomic background
determined by N-adapted frames (\ref{nadif}) for a generic off-diagonal
solution (\ref{qelgenofd}) with limits $\widehat{\mathbf{D}}\rightarrow
\nabla $ and $\mathbf{g}_{\alpha \beta }\rightarrow \mathbf{g}_{\alpha \beta
}(t,a(t),\widehat{h}_{3}(t),\check{\Phi}(t),\eta _{i}(t)).$ With respect to
such frames, the nonholonomic FLRW equations are similar to those found in
section III B of \cite{odgom} (see the second paper for details on methods
of constructing solutions and speculations on the problem of matter
instability).\footnote{%
In section III A of that work, a model with $G(T)=0$ was investigated in
detail. The conclusion was that in order to elaborate a realistic evolution
it is necessary to consider nontrivial values for $G(T).$ In nonholonomic
variables, such term $\widehat{\mathbf{G}}(\widehat{\mathbf{T}})$ allows to
encode $f(R)$ modified theories and related into certain off-diagonal
configurations in GR, which simplifies the solution of the problem of matter
instability (see subsection \ref{ssminst}).}

The effective function $a(t)$ defines in our case off-diagonal cosmological
evolution scenarios which are different from those where $\mathring{a}(t)$
stands for a standard diagonal FLRW cosmology. For $H:=a^{\diamond }/a,$ $\
^{1}\widehat{\mathbf{G}}:=d\widehat{\mathbf{G}}/d\widehat{\mathbf{T}}$ and $%
\ ^{1}\widehat{\mathbf{F}}:=d\widehat{\mathbf{F}}/d\widehat{\mathbf{P}},$ we
have%
\begin{eqnarray}
3H^{2}+\frac{1}{2}\left[ \widehat{\mathbf{f}}+\widehat{\mathbf{G}}%
-3(3H^{2}-H^{\diamond })\ \rho \ ^{1}\widehat{\mathbf{F}}\right] -\rho
(\kappa ^{2}-\ ^{1}\widehat{\mathbf{G}}) &=&0,  \label{ceq} \\
-3H^{2}-2H^{\diamond }-\frac{1}{2}[\widehat{\mathbf{f}}+\widehat{\mathbf{G}}%
-\left( \rho \ ^{1}\widehat{\mathbf{F}}\right) ^{\diamond \diamond
}-4H\left( \rho \ ^{1}\widehat{\mathbf{F}}\right) ^{\diamond }-\left(
3H^{2}+H^{\diamond }\right) \ \rho \ ^{1}\widehat{\mathbf{F}}] &=&0.  \notag
\end{eqnarray}%
An observer is here in a nonholonomic basis determined by $%
N_{i}^{a}=\{n_{i},w_{i}(t)\}$ for a nontrivial off-diagonal vacuum with
effective polarizations $\eta _{\alpha }(t)$, and can test cosmological
scenarios in terms of the redshift $1+z=a^{-1}(t)$ for $P=P(z)$ and $T=T(z),$
with a new \textquotedblleft shift\textquotedblright\ derivative when (for
instance, for a function $s(t)$) $s^{\diamond }=-(1+z)H\partial _{z}.$

The system of two equations (\ref{ceq}) simplifies by extending it to a set
of three equations for four unknown functions $\{\widehat{\mathbf{f}}(z),%
\widehat{\mathbf{G}}(z),\rho (z),\varsigma (z)\}$ with a new variable $%
\varsigma (z):=\ \rho \ ^{1}\widehat{\mathbf{F}},$
\begin{eqnarray}
3H^{2}+\frac{1}{2}[\widehat{\mathbf{f}}(z)+\widehat{\mathbf{G}}(z)]-\frac{3}{%
2}[3H^{2}-(1+z)H(\partial _{z}H)]\ \varsigma (z) \frac{3}{2}%
H^{2}(1+z)\partial _{z}\varsigma (z)-\kappa ^{2}\rho (z)&=&0,  \notag \\
-3H^{2}+(1+z)H(\partial _{z}H)-\frac{1}{2}\{\widehat{\mathbf{f}}(z)+\widehat{%
\mathbf{G}}(z)-[3H^{2}-(1+z)H(\partial _{z}H)]\varsigma (z) &&  \notag \\
+[3(1+z)H^{2}-(1+z)H(\partial _{z}H)]\partial _{z}\varsigma
(z)+(1+z)^{2}\partial _{zz}^{2}\varsigma (z)\}&=&0,  \notag \\
(\partial _{z}\ ^{1}\widehat{\mathbf{F}})\ \varsigma (z)-\rho (z)\ (\partial
_{z}\ \widehat{\mathbf{f}})&=&0.  \label{ceq1}
\end{eqnarray}%
Here, by re-scaling the generating function, we have fixed the condition $%
\partial _{z}\ ^{1}\widehat{\mathbf{G}}(z)=0$. Such a nontrivial term must
be considered if one wants to transform $\widehat{\mathbf{f}}$ into a
standard theory $f(R,T,R_{\alpha \beta }T^{\alpha \beta })$. The functional $%
\widehat{\mathbf{G}}(\widehat{\mathbf{T}})$, in both holonomic and
nonholonomic forms, encodes a new degree of freedom for the evolution of the
energy-density of type $\rho =\rho _{0}a^{-3(1+\varpi )}=\rho
_{0}(1+z)a^{3(1+\varpi )}$, which is taken for the dust matter approximation
$\varpi $ when the evolution reduces to $\rho \sim (1+z)^{3}.$ For the
assumption that such an evolution can be considered with respect to
N-adapted frames, the solutions of (\ref{ceq1}) are determined by data $\{%
\widehat{\mathbf{f}}(z),\widehat{\mathbf{G}}(z),\varsigma (z)\}$ by
replacing the second and third equations into the first one and obtaining a
single fourth-order equation for $\widehat{\mathbf{f}}(z).$

The reconstruction procedure is restricted to fluids without pressure when
such approximation is considered locally with N-adapted frames and the
expressions (\ref{auxaa}) for $(\mathring{a},\mathring{H},\mathring{\rho})$
are re-defined in terms of $(a,H,\rho )$; data are written with a script
\textquotedblleft 0\textquotedblright\ if $z=z_{0},$ with $\xi =\kappa
^{2}\rho _{0}/3H_{0}^{2}$. One should not confused, e.g., $\mathring{H}$ and
$H_{0}$, because these values are computed for \textit{different} FLRW
solutions, with $\mathring{a}(z)$ determined for a diagonal configuration
and $a(z)$ for an off-diagonal one, respectively. We can express  $\widehat{%
\mathbf{T}}=\widehat{\mathbf{T}}_{\ \alpha }^{\alpha }=-\xi \frac{3H_{0}^{2}%
}{\kappa ^{2}}(1+z)^{3}$ and $\widehat{\mathbf{P}}=\widehat{\mathbf{R}}%
_{\alpha \beta \ }\widehat{\mathbf{T}}_{\ }^{\alpha \beta }=-3\xi \frac{%
3H_{0}^{2}}{\kappa ^{2}}(1+z)^{3}[H^{2}-(1+z)H(\partial _{z}H)]$.  Following
the approach outlined in Sect.~IIIB of \cite{odgom}, we introduce the
parameterizations
\begin{equation}
\widehat{\mathbf{F}}(\widehat{\mathbf{P}})=H_{0}^{2}\mathbf{\check{F}}(%
\mathbf{\check{P}})\mbox{ and }\widehat{\mathbf{G}}(\widehat{\mathbf{T}}%
)=H_{0}^{2}\mathbf{\check{G}}(\mathbf{\check{T}}),  \label{aux9}
\end{equation}%
where $\mathbf{\check{P}=}\widehat{\mathbf{P}}/P_{0}$ and $\mathbf{\check{T}=%
}$ $\widehat{\mathbf{T}}/T_{0},$ for $P_{0}=-9H_{0}^{4}\xi /\kappa ^{2}$ and
$T_{0}=-3H_{0}^{2}\xi /\kappa ^{2}.$ In correspondingly N-adapted variables,
the off-diagonal cosmological solutions can be associated with a class of de
Sitter (dS) solutions with effective cosmological constant $\check{\Lambda}$%
, where $H(z)=\check{H}_{0}$ results in $\mathbf{\check{P}}=\mathbf{\check{T}%
}=(1+z)^{3}$. In these variables, the solutions of (\ref{ceq1}) can be
written as
\begin{eqnarray}
\mathbf{\check{F}} &=&c_{1}\mathbf{\check{P}}^{b_{1}}+\mathbf{\check{P}}%
^{b_{2}/3}[c_{2}\cos (\frac{b_{3}}{3}\ln \mathbf{\check{P}})+c_{3}\sin (%
\frac{b_{3}}{3}\ln \mathbf{\check{P}})]+c_{4}+3\xi \mathbf{\check{P},}
\label{aux8} \\
\mathbf{\check{G}} &=&\tilde{c}_{1}\mathbf{\check{T}}^{b_{1}}+\mathbf{\check{%
T}}^{b_{2}/3}[\tilde{c}_{2}\cos (\frac{b_{3}}{3}\ln \mathbf{\check{T}})+%
\tilde{c}_{3}\sin (\frac{b_{3}}{3}\ln \mathbf{\check{P}})]+\tilde{c}%
_{4}-3\xi \mathbf{\check{T},}  \notag
\end{eqnarray}%
being the constants $b_{1}=-1.327,b_{2}=3.414$ and $b_{3}=1.38.$ The values $%
c_{1},c_{2},c_{3}$ and $c_{4}$ are integration constants, and the second set
of constants $\tilde{c}_{1},\tilde{c}_{2},\tilde{c}_{3}$ and $\tilde{c}_{4}$
can be expressed via such integration constants, and $b_{1},b_{2}$ and $%
b_{3}.$ We omit explicit formulas because for general solutions they can be
included in certain generating or integration functions for the modified
gravity equations and ultimately related to real observation data for the
associated cosmological models.

For off-diagonal configurations, the $\widehat{\mathbf{f}}(\widehat{\mathbf{R%
}},\widehat{\mathbf{T}},$ $\widehat{\mathbf{R}}_{\alpha \beta }\widehat{%
\mathbf{T}}^{\alpha \beta })$ gravity positively allows for dS solutions in
presence of non-constant fluids, not only due to the term $\widehat{\mathbf{P%
}}=\widehat{\mathbf{R}}_{\alpha \beta }\widehat{\mathbf{T}}^{\alpha \beta }$
in (\ref{mgts}), and respective gravitational field and cosmological
equations. This is possible also because of the off-diagonal nonlinear
gravitational interactions in the effective gravitational models. It should
be emphasized that the reconstruction procedure elaborated in \cite{odgom},
see also references therein, can be extended to more general classes of
modified gravity theories, to Finsler like theories and the ensuing
cosmological models \cite{voffdmgt}. Introducing (\ref{aux8}) and (\ref{aux9}%
) into (\ref{aux10}), we reconstruct a function $\widehat{\mathbf{f}}=%
\widehat{\mathbf{R}}+\widehat{\mathbf{F}}(\widehat{\mathbf{P}})+\widehat{%
\mathbf{G}}(\widehat{\mathbf{T}}).$ As a result, we can associate an
effective matter source $\ \widehat{\mathbf{\Upsilon }}$, which allows the
definition of a corresponding generating function $\check{\Phi}$  (see also $\Phi $ and (\ref{aux2})). Finally, we can reconstruct an off-diagonal cosmological solution with nonholonomically induced torsion of
type (\ref{odfrlwtors}) or to model a similar cosmological metric for LC
configurations (\ref{qelgen}) (equivalently, (\ref{qelgenofd})).

\subsection{How $f$--gravity and cosmology are encoded in nonholonomic
configurations?}

It is well known that any FLRW cosmology can be realized in a specific $f(R)$
gravity (see Ref.~\cite{odintsplb} and, for further generalizations, \cite%
{odgom}).\footnote{%
We use a system of notations different from that article;\ here, e.g., $N$
in used for the N-connection and we work with nonholonomic geometric objects.%
} In this subsection we analyze two examples of reconstruction of $f(R)$%
-gravities where the ''e-folding'' variable $\zeta :=\ln a/a_{0}=-\ln (1+z) $
is used instead of the cosmological time $t$ and in related nonholonomic
off-diagonal deformations. For such models, we consider $\widehat{\mathbf{f}}%
=\widehat{\mathbf{f}}(\widehat{\mathbf{R}})$ in (\ref{mgts}), use $\
\widehat{\mathbf{\Upsilon }}(x^{i},\zeta )=\Lambda /\ ^{1}\widehat{\mathbf{f}%
}+\widehat{\mathbf{f}}/2\ ^{1}\widehat{\mathbf{f}}$ instead of (\ref%
{effsource}), which can be parameterized with dependencies on $(x^{i},\zeta )
$ (in particular, only on $\zeta $), $\check{\Phi}^{2}=\check{\Lambda}^{-1}[%
\widehat{\Phi }^{2}|\ \widehat{\mathbf{\Upsilon }}|+\int d\zeta \ \widehat{%
\Phi }^{2}\partial _{\zeta }|\ \widehat{\mathbf{\Upsilon }}|],$ when $%
\partial _{\zeta }=\partial /\partial \zeta $ with $s^{\diamond }=H\partial
_{\zeta }s$ for any function $s.$ The matter energy density $\rho $ is taken
as in (\ref{ceq1}).

We restrict ourselves to N-adapted frames (\ref{nadif}), determined by an
off-diagonal cosmological solution of the (modified) gravitational field
equations, and can repeat all computations leading to Eqs.~(2)-(7) in \cite%
{odintsplb} and prove that a modified gravity with $\widehat{\mathbf{f}}(%
\widehat{\mathbf{R}})$ realizes the FLRW cosmological model. Such solutions
depend on the above source type $\widehat{\mathbf{\Upsilon }}(x^{i},\zeta )$
and generating function $\check{\Phi}(x^{i},\zeta )$; also the nonholonomic
background can be modeled to be nonhomogeneous (via $w_{i}$ and $n_{i}$
depending respectively on $x^{i}$ and $\zeta ,$ or only on $\zeta ).$ The
off-diagonal analog of the field equation corresponding to the first FLRW
equation is {\small \
\begin{equation*}
\widehat{\mathbf{f}}(\widehat{\mathbf{R}})=(H^{2}+H\ \partial _{\zeta
}H)\partial _{\zeta }[\widehat{\mathbf{f}}(\widehat{\mathbf{R}})]-36H^{2}%
\left[ 4H+(\partial _{\zeta }H)^{2}+H\partial _{\zeta \zeta }^{2}H\right]
\partial _{\zeta \zeta }^{2}[\widehat{\mathbf{f}}(\widehat{\mathbf{R}}%
\mathbf{)]+}\kappa ^{2}\rho .
\end{equation*}%
} In terms of an effective quadratic Hubble rate, $q(\zeta ):=H^{2}(\zeta ),$
and considering that $\zeta =\zeta (\widehat{\mathbf{R}})$ for certain
parameterizations, this equation yields {\small
\begin{equation}
\widehat{\mathbf{f}}(\widehat{\mathbf{R}})=-18q(\zeta (\widehat{\mathbf{R}}%
))[\partial _{\zeta \zeta }^{2}q(\zeta (\widehat{\mathbf{R}}))+4\partial
_{\zeta }q(\zeta (\widehat{\mathbf{R}}))]\frac{d^{2}\widehat{\mathbf{f}}(%
\widehat{\mathbf{R}})}{d\widehat{\mathbf{R}}^{2}}+6[q(\zeta (\widehat{%
\mathbf{R}}))+\frac{1}{2}\partial _{\zeta }q(\zeta (\widehat{\mathbf{R}}))]
\frac{d\widehat{\mathbf{f}}(\widehat{\mathbf{R}})}{d\widehat{\mathbf{R}}}%
+2\rho _{0}a_{0}^{-3(1+\varpi )}a^{-3(1+\varpi )\zeta (\widehat{\mathbf{R}}%
)}.  \label{f1gen}
\end{equation}%
} We can construct an off-diagonal cosmological model with metrics of type (%
\ref{odfrlwtors}) and nonholonomically induced torsion (when $t\rightarrow
\zeta )$ if a solution $\widehat{\mathbf{f}}(\widehat{\mathbf{R}})$ is used
for computing $\widehat{\mathbf{\Upsilon }}$ and $\check{\Phi}.$ Modeling
such nonlinear systems we can consider solutions of the field equations for
an effective (nonholonomic) Einstein space $\mathbf{\check{R}}_{\ \beta
}^{\alpha }=\check{\Lambda}\delta _{\ \beta }^{\alpha }$, when certain terms
of type $d\widehat{\mathbf{f}}(\mathbf{\check{R}})/d\mathbf{\check{R}}$ and
higher derivatives vanish for a functional dependence $\widehat{\mathbf{f}}(%
\check{\Lambda})$ with $\partial _{\zeta }\check{\Lambda}=0.$ The
nonholonomic cosmological evolution is determined by off-diagonal
coefficients of the metrics and by certain non-explicit relations for the
functionals variables, like $q(\zeta (\widehat{\mathbf{R}}(\check{\Lambda}%
))) $ and (effective/modified) matter sources transform as $\ \widehat{%
\mathbf{\Upsilon }}$ (\ref{dsours1}) $\rightarrow $ $\check{\Lambda}$ (\ref%
{dsours2}).

LC-configurations can be modeled by off-diagonal cosmological metrics of
type (\ref{qelgen}) when the zero torsion conditions (\ref{lccondb}) are
satisfied. We obtain a standard expression (see \cite{odintsplb}) for the
curvature of $\nabla ,$
\begin{equation}
R=3\partial _{\zeta }q(\zeta )+12q(\zeta ),  \label{cscurv}
\end{equation}
if the polarization or generating functions for (\ref{qelgen}) and the
solutions of (\ref{f1gen}) are taken for diagonal configurations.

We here provide an example of reconstruction of models of $f(R)$ gravity and
nonholonomically deformed GR when both reproduce the ${\Lambda}$CDM era. For
simplicity, we do not consider a real matter source (if such a source
exists, it can be easily encoded into a nontrivial vacuum structure with
generic off-diagonal contributions).

With respect to correspondingly N-adapted frames and for $a(\zeta )$ and $%
H(\zeta )$ determined by an off-diagonal solution (\ref{odfrlwtors}), with
nonholonomically induced torsion, or (\ref{qelgenofd}), for
LC-configurations, the FLRW equation for ${\Lambda}$CDM cosmology is given
by
\begin{equation}
3\kappa ^{-2}H^{2}=3\kappa ^{-2}H_{0}^{2}+\rho _{0}a^{-3}=3\kappa
^{-2}H_{0}^{2}+\rho _{0}a_{0}^{-3}e^{-3\zeta }.  \label{aux11}
\end{equation}%
This equation looks similar to the one for Einstein gravity for diagonal
configurations but contains values determined, in general, for other classes
of models with off-diagonal interactions. Thus, $H_{0} $ and $\rho _{0}$ are
fixed to be certain constant values, after the coefficients of off-diagonal
solutions are found, and for an approximation were the dependencies on $%
(x^{i},\zeta )$ are changed into dependencies on $\zeta $ (via a
corresponding re-definition of the generating functions and the effective
sources). We can relate the first term on the rhs to an effective
cosmological constant $\Lambda $ (\ref{dsours2}), which in our approach
appears via a re-definition (\ref{aux2}). The second term in the formula
describes, in general, an inhomogeneous distribution of cold dark mater
(CDM) with respect to N-adapted frames. In order to keep the similarity with
the diagonalizable cosmological models in GR we can choose these integration
constants for $\Lambda =12H_{0}^{2}$ to survive in the limit $%
w_{i},n_{i}\rightarrow 0.$ It should be noted that such limit must be
computed for ``nonlinear'' nonholonomic constraints via generating functions
and effective sources.

Using (\ref{aux11}), the effective quadratic Hubble rate and the modified
scalar curvature, $\widehat{\mathbf{R}}$, are computed to be, respectively,
$q(\zeta ) := H_{0}^{2}+\kappa ^{2}\rho _{0}a_{0}^{-3}e^{-3\zeta }$ and $%
\widehat{\mathbf{R}} = 3\partial _{\zeta }q(\zeta )+12q(\zeta
)=12H_{0}^{2}+\kappa ^{2}\rho _{0}a_{0}^{-3}e^{-3\zeta }$. These functional
formulas can be used for the dependencies on $\widehat{\mathbf{R}}$ if a
necessary re-definition of the generation functions, or an approximation $%
(x^{i},\zeta )\rightarrow \zeta $ is performed. Expressing  $3\zeta =-\ln
[\kappa ^{-2}\rho _{0}^{-1}a_{0}^{3}(\widehat{\mathbf{R}}-12H_{0}^{2})]$ and
$X:=-3+\widehat{\mathbf{R}}/3H_{0}^{2}$, we obtain from Eq.~(\ref{f1gen})%
\begin{equation}
X(1-X)\frac{d^{2}\widehat{\mathbf{f}}}{dX^{2}}+[\chi _{3}-(\chi _{1}+\chi
_{2}+1)X]\frac{d\widehat{\mathbf{f}}}{dX}-\chi _{1}\chi _{2}\widehat{\mathbf{%
f}}=0,  \label{gauss}
\end{equation}%
for certain constants, for which $\chi _{1}+\chi _{2}=\chi _{1}\chi
_{2}=-1/6 $ and $\chi _{3}=-1/2.$ The solutions of this equation with
constant coefficients and for $R$ (\ref{cscurv}) were found in \cite%
{odintsplb} as Gauss hypergeometric function, denoted there by $\widehat{%
\mathbf{f}}=F(X):=F(\chi _{1},\chi _{2},\chi _{3};X),$ as \  $F(X)=AF(\chi
_{1},\chi _{2},\chi _{3};X)+BX^{1-\chi _{3}}F(\chi _{1}-\chi _{3}+1,\chi
_{2}-\chi _{3}+1,2-\chi _{3};X)$  (for some constants $A$ and $B $). This
provides a proof of the statement that $f(R)$ gravity can indeed describe ${%
\Lambda}$CDM scenarios without the need of an effective cosmological
constant. Working with auxiliary connections of the type $\widehat{\mathbf{D}%
},$ we can generalize the constructions to off-diagonal configurations and
various classes of modified gravity theories, where $A,B$ and $\chi
_{1},\chi _{2},\chi _{3}$ are appropriate functions of the $h $ coordinates.
For instance, reconstruction procedures for Finsler like theories and
cosmology models on tangent and Lorentz bundles, and bi-metric/massive
gravity models are given in \cite{voffdmgt,massgr}.

Having chosen $\widehat{\mathbf{f}}=F(X)$ for a modified gravity, we can go
further and mimic an off-diagonal configuration when $\widehat{\mathbf{f}}=%
\widehat{\mathbf{f}}(\widehat{\mathbf{R}})$ is introduced in (\ref{mgts})
and the source $\ \widehat{\mathbf{\Upsilon }}(x^{i},\zeta )=\Lambda /\ ^{1}%
\widehat{\mathbf{f}}+\widehat{\mathbf{f}}/2\ ^{1}\widehat{\mathbf{f}}$ is
considered instead of (\ref{effsource}) and (\ref{aux2b}) for $\check{\Phi}%
^{2}=\check{\Lambda}^{-1}[\widehat{\Phi }^{2}|\ \widehat{\mathbf{\Upsilon }}%
|+\int d\zeta \ \widehat{\Phi }^{2}\partial _{\zeta }|\ \widehat{\mathbf{%
\Upsilon }}|].$ Nevertheless, recovering nonhomogeneous modified
cosmological models cannot be completed for general re-parameterized
dependencies on $(x^{i},\zeta )$ (in particular, only on $\zeta $). This
distinguishes explicitly the modified gravity theories of type $f(R)$ from
those generated by nonholonomic deformations. For certain homogeneity
conditions, we can state an equivalence of some classes of gravities and
cosmological models, or analyze their alternative physical implications. But
a complete recovering is only possible if all generating and integration
functions and the effective sources are correlated with certain observable
cosmological effects and further approximations and re-definitions in terms
of constant parameters and functionals depending on a time-like coordinate
can be effectively performed.

The AFDM allows to reconstruct off-diagonal configurations modeling $f(R)$
gravity and cosmology with non-phantom or phantom matter in GR. With respect
to N-adapted frames in an off-diagonal (modified, or not) gravitational
background, the FLRW equations can be written as
\begin{equation}
3\kappa ^{-2}H^{2}=\ _{s}\rho (x^{k})a^{-c(x^{k})}+\ _{p}\rho
(x^{k})a^{c(x^{k})},  \label{aux11a}
\end{equation}%
where $a(x^{k},\zeta )$ and $H(x^{k},\zeta )$ are determined by a solution (%
\ref{odfrlwtors}), or (\ref{qelgenofd}). For re-parameterizations or
approximations with $(x^{i},\zeta )\rightarrow \zeta ,$ we assume that the
positive functions$\ _{s}\rho (x^{k}),\ _{p}\rho (x^{k})$ and $c(x^{k})$ can
be considered. The first term on the rhs dominates for small $a$ in the
early universe, as in GR with non-phantom matter described by an EoS
parameter $w=-1+c/3>-1.$ Introducing $q(x^{k},\zeta ):=H^{2}(x^{k},\zeta )$
and the respective functionals $\ _{s}q:=\frac{\kappa ^{2}}{3}\ _{s}\rho
a_{0}^{-c}$ and $\ _{p}q:=\frac{\kappa ^{2}}{3}\ _{p}\rho a_{0}^{c},$ for $%
q=\ _{s}qe^{-c\zeta }+$ $\ _{p}qe^{c\zeta },$ in $\widehat{\mathbf{R}}%
=3\partial _{\zeta }q(\zeta )+12q(\zeta ),$ we find
\begin{equation}
e^{c\zeta }=\left\{
\begin{array}{cc}
\lbrack \widehat{\mathbf{R}}\pm \sqrt{\widehat{\mathbf{R}}^{2}-4(144-9c^{2})}%
]/6(4+c), & \mbox{ for }c\neq 4; \\
\widehat{\mathbf{R}}/24, & \mbox{ for }c=4.%
\end{array}%
\right.  \label{aux12}
\end{equation}
The non-phantom matter may correspond to the case $c=4$ in (\ref{aux12}),
including radiation with $w=1/3.$ Eq.~(\ref{aux11a}) transform into a
functional equation on $Y$ determined by changing the functional variable $%
\widehat{\mathbf{R}}^{2}=-576\ _{s}q\ \ _{p}q\ Y,$ $4Y(1-Y)\frac{d^{2}%
\widehat{\mathbf{f}}}{dY^{2}}+(3+Y)\frac{d\widehat{\mathbf{f}}}{dY}-2%
\widehat{\mathbf{f}}=0$. This is again a functional variant (if we consider
dependencies on $x^{k}$) of the generating Gauss' hypergeometric function,
similarly to (\ref{gauss}), which can be solved in explicit form.

For the case $c\neq 4$ in (\ref{aux12}), we come to models with phantom-like
dominant components. A similar procedure as for deriving Eqs.~(22) and (23)
in \cite{odintsplb}, results in a functional generalization of the Euler
equation, namely  $\widehat{\mathbf{R}}^{2}\frac{d^{2}\widehat{\mathbf{f}}(%
\widehat{\mathbf{R}})}{d\widehat{\mathbf{R}}^{2}}+A\widehat{\mathbf{R}}\frac{%
d\widehat{\mathbf{f}}(\widehat{\mathbf{R}})}{d\widehat{\mathbf{R}}}+B%
\widehat{\mathbf{f}}(\widehat{\mathbf{R}})=0$, for some coefficients $%
A=-H_{0}(1+H_{0})$ and $B=(1+2H_{0})/2,$ for $H_{0}=1/3(1+\ _{ph}w)$. Here
we consider, for simplicity, homogenous limits and approximations $H^{2}(t)=%
\frac{\kappa ^{2}}{3}_{ph}\rho $ for the phantom EoS fluid-like states, $%
_{ph}p=\ _{ph}w\ _{ph}\rho ,$ with $\ _{ph}w<-1.$ In both cases, with a
trivial or a nontrivial nonholonomically induced torsion, there are
solutions of the nonholonomic Euler equations above which can be expressed
in the form $\widehat{\mathbf{f}}(\widehat{\mathbf{R}})=C_{+}\widehat{%
\mathbf{R}}^{m_{+}}+C_{-}\widehat{\mathbf{R}}^{m_{-}}$, for some integration
constants $C_{\pm }$ and $2m_{\pm }=1-A\pm \sqrt{(1-A)^{2}-4B}.$ This
reproduces with respect to N-adapted frames the phantom dark energy
cosmology with a behavior of the type $a(t)=a_{0}(t_{s}-t)^{-H_{0}},$ where $%
t_{s} $ is the so-called Rip time. If the generating functions for the
off-diagonal cosmological solutions are chosen in a way such that the
N-connection coefficients $w_{i}$ and $n_{i}$ transform to zero, the
solutions describe universes which end at a Big Rip singularity during $%
t_{s}. $ Additionally to the former result that in the $f(R)$ theory no
phantom fluid is needed, we conclude that for off-diagonal configurations we
can effectively model such locally anisotropic cosmological configurations.

One can encode and effectively model various types of cosmological solutions
for modified gravity theories with $f(R)$ and/or $f(R,T,R_{\alpha \beta
}T^{\alpha \beta })$ functionals and their nonholonomic deformations. The
cosmological reconstruction procedures can be elaborated for various types
of viable modified gravity which may pass, or not, local gravitational tests
and explain observational data for accelerating cosmology, dark energy and
dark matter interactions \cite{odgom,odintsplb,massgr,voffdmgt}.
Nevertheless, these theories exhibit certain specific problems as
non-conservation of the energy-momentum tensors for the effective or
physical matter fields.

In explicit form, we explain how the \textquotedblleft
non-conservation\textquotedblright\ problem can be solved for off-diagonal
solutions with one Killing symmetry in the framework of $f(R,T)$ theories
generalizing certain constructions from \cite{alvar}. Following a similar
procedure as in Sect.~II of that work, but using the operator $\widehat{%
\mathbf{D}}$ instead of $\nabla ,$ for $\widehat{\mathbf{f}}=\widehat{%
\mathbf{f}}(\widehat{\mathbf{R}},\widehat{\mathbf{T}})$, and considering an
N-adapted parametrization of the effective source $\widehat{\mathbf{\Upsilon
}}=const$ $,$ we prove that
\begin{equation}
(1+\frac{\kappa ^{2}}{\ ^{2}\widehat{\mathbf{f}}})\widehat{\mathbf{D}}%
^{\alpha }\widehat{\mathbf{T}}_{\alpha \beta }=\frac{1}{2}\mathbf{g}_{\alpha
\beta }\widehat{\mathbf{D}}^{\alpha }\widehat{\mathbf{T}}-(\widehat{\mathbf{T%
}}_{\alpha \beta }+\widehat{\Theta }_{\alpha \beta })\widehat{\mathbf{D}}%
^{\alpha }\ln (\ ^{2}\widehat{\mathbf{f}})-\widehat{\mathbf{D}}^{\alpha }%
\widehat{\Theta }_{\alpha \beta }.  \label{divemt}
\end{equation}%
In these equations the values $\ ^{2}\widehat{\mathbf{f}}:=\partial \widehat{%
\mathbf{f}}/\partial \widehat{\mathbf{T}}$ and $\widehat{\Theta }_{\alpha
\beta }=-2\widehat{\mathbf{T}}_{\alpha \beta }-p\mathbf{g}_{\alpha \beta }$
are used, with an energy-momentum tensor (\ref{dsourc}) for nonholonomic
flows of a perfect fluid. In general, $\widehat{\mathbf{D}}^{\alpha }%
\widehat{\mathbf{T}}_{\alpha \beta }\neq 0$ even for nonholonomic
deformations of GR. Nevertheless, we can consider a subclass of off-diagonal
configurations in $\widehat{\mathbf{f}}(\widehat{\mathbf{R}},\widehat{%
\mathbf{T}})$ gravity when $\ \widehat{\mathbf{\Upsilon }}$ (\ref{dsours1}) $%
\rightarrow $ $\check{\Lambda}$ (\ref{dsours2}) and $\check{\Phi}^{2}=\check{%
\Lambda}^{-1}[\widehat{\Phi }^{2}|\ \widehat{\mathbf{\Upsilon }}|+\int
d\zeta \ \widehat{\Phi }^{2}\partial _{\zeta }|\ \widehat{\mathbf{\Upsilon }}%
|]$ result in $\widehat{\mathbf{f}}\rightarrow \mathbf{\check{f}=\check{R}}$
and effective $\mathbf{\check{R}}_{\ \beta }^{\alpha }=\check{\Lambda}\delta
_{\ \beta }^{\alpha }$ which admit LC-solutions with zero torsion. For such
nonholonomic distributions with $\widehat{\mathbf{D}}\rightarrow \nabla ,$ $%
\widehat{\mathbf{D}}^{\alpha }\widehat{\mathbf{T}}_{\alpha \beta
}\rightarrow \check{\nabla}\check{\Lambda}=0$ and all terms on the lhs of (%
\ref{divemt}) vanish, because they are nonholonomically equivalent to
functionals o f the effective cosmological constant $\check{\Lambda}.$ Such
conditions are satisfied in correspondingly N-adapted frames and for
canonical d-connections. The equations (\ref{divemt}) generalize to
nonholonomic forms the similar ones derived for the Levi-Civita connection $%
\nabla $ (see Eq.~(10) in Ref.~\cite{alvar}).

\subsection{Nonholonomic constraints and matter instability}

\label{ssminst} There is another serious problem in modified gravities which
is related to possible matter instabilities related to modifications of the
gravitational actions. Even tiny modifications of GR may make the new model
to posses unstable interior solutions (see, e.g., \cite{minst}). It was
demonstrated however that there are viable $f(R)$ theories (with
appropriated choices of the functional) where such instabilities may not
occur \cite{revfmod,stabodin}. The issue of instability and stabilization via additional nonholonomic constraints will be studied in our further works. In this section, we  speculate how the AFDM can be applied to stability analysis in  more general $f(R,T,R_{\alpha \beta }T^{\alpha
\beta })$ theories. The corresponding field equations are very difficult to
solve even in a linear approximation \cite{odgom}, if we work in coordinate
frames and with general functionals. In the nonholonomic variable formalism, the
gravitational field equations in modified gravity theories posses the
decoupling property exhibited above, which allows to encode $f(R,...)$%
-modifications into off-diagonal nonholonomic configurations for the
effective Einstein manifolds.

For a stability analysis, the trace equations where (\ref{mgtfe}) are
multiplied by $\mathbf{g}^{\mu \nu }$ are to be considered, namely%
\begin{equation}
-2\widehat{\mathbf{f}}+(\widehat{\mathbf{R}}\ +3\widehat{\mathbf{D}}^{\mu }%
\widehat{\mathbf{D}}_{\mu })\ ^{1}\widehat{\mathbf{f}}+(\widehat{\mathbf{T}}+%
\mathbf{\Theta })\ ^{2}\widehat{\mathbf{f}}+(\frac{1}{2}\widehat{\mathbf{D}}%
^{\mu }\widehat{\mathbf{D}}_{\mu }\widehat{\mathbf{T}}+\widehat{\mathbf{D}}%
_{\mu }\widehat{\mathbf{D}}_{\nu }\widehat{\mathbf{T}}^{\mu \nu }+\mathbf{%
\Xi )}\ ^{3}\widehat{\mathbf{f}}=\kappa ^{2}\ \widehat{\mathbf{T}},
\label{lineardef}
\end{equation}
where $\ ^{1}\widehat{\mathbf{f}}:=\partial \widehat{\mathbf{f}}/\partial
\widehat{\mathbf{R}},$ $\ ^{2}\widehat{\mathbf{f}}:=\partial \widehat{%
\mathbf{f}}/\partial \widehat{\mathbf{T}}$ and $\ \ ^{3}\widehat{\mathbf{f}}%
:=\partial \widehat{\mathbf{f}}/\partial \widehat{\mathbf{P}},$ when $%
\widehat{\mathbf{P}}=\widehat{\mathbf{R}}_{\alpha \beta }\widehat{\mathbf{T}}%
^{\alpha \beta }.$ Let us envisage a trace configuration in the interior of
a celestial body, when $\widehat{\mathbf{T}}=\widehat{\mathbf{T}}_{0}$ and $%
-2\widehat{\mathbf{f}}+\widehat{\mathbf{R}}_{0}\ (\ ^{1}\widehat{\mathbf{f}}%
)=\kappa ^{2}\ \widehat{\mathbf{T}}_{0}.$ Imposing nonholonomic constraints,
we parameterize a LC-configuration in GR and model an interior solution in
the presence of some gravitational objects (for instance, the Sun or the
Earth). The $f$-modifications (in general with strong coupling for the
curvature and the energy-momentum tensor) may result in a worsening of the
stability problems and may prevent $\widehat{\mathbf{T}}_{0}$ to be a
solution of any suitable background equation. It is difficult to find
solutions of (\ref{lineardef}) even for very much simplified cases in the
nonlinear situation if we work in coordinate frames for the connection $%
\widehat{\mathbf{D}}=\nabla .$

A rigorous study of the problem of matter instability for $f(R)$ and more
generally $f(R,T,R_{\alpha \beta }T^{\alpha \beta })$ gravities, for certain
illustrative cases when $\ ^{1}\widehat{\mathbf{f}}=R$, and for restrictive
conditions where there is a qualitative description via additional
functionals on $T$ and $P$ shows that the appearance of a large instability
can actually be avoided. Using the AFDM, we
can consider modified gravity theories with $f$-modifications which are
effectively described by $\mathbf{\check{R}}_{\ \beta }^{\alpha }=\check{%
\Lambda}\delta _{\ \beta }^{\alpha }$ when the modifications are encoded
into polarization functions and N-coefficients. For models generated by $\
\widehat{\mathbf{f}}(\widehat{\mathbf{R}},\widehat{\mathbf{T}},\widehat{%
\mathbf{R}}_{\alpha \beta }\widehat{\mathbf{T}}^{\alpha \beta })=\widehat{%
\mathbf{f}}_{1}(\widehat{\mathbf{R}})+\widehat{\mathbf{F}}(\widehat{\mathbf{P%
}})+\widehat{\mathbf{G}}(\widehat{\mathbf{T}}),$ we take a constant interior
solution with $\widehat{\mathbf{T}}_{0}=const$ and $\widehat{\mathbf{P}}%
_{0}=const,$ and denote by $\widehat{\mathbf{f}}_{1}^{(s)}:=\partial ^{s}%
\widehat{\mathbf{f}}_{1}/\partial \widehat{\mathbf{R}}^{s}$ and $\widehat{%
\mathbf{F}}^{(s)}:=\partial ^{s}\widehat{\mathbf{F}}/\partial \widehat{%
\mathbf{P}}^{s}$ for $s=1,2,...$ We can repeat, with respect to the N-frames
(\ref{nadif}), the computations presented in detail for Eqs.~(45)-(48) in
\cite{odgom} (see also references therein), and prove that Eqs.~(\ref%
{lineardef}) for linear pertubations can be written in the form  $$\lbrack
\mathbf{\check{D}}^{\mu }\mathbf{\check{D}}_{\mu }+2\frac{\mathbf{\check{T}}%
_{0}^{\mu \nu }}{\mathbf{\check{T}}_{0}}\mathbf{\check{D}}_{\mu }\mathbf{%
\check{D}}_{\nu }+2\frac{\mathbf{\Xi }_{0}}{\mathbf{\check{T}}_{0}}+4\frac{%
\mathbf{\check{P}}_{0}}{\mathbf{\check{T}}_{0}}\frac{\widehat{\mathbf{f}}%
_{1}^{(1)}}{\widehat{\mathbf{F}}^{(2)}}]\ \delta \mathbf{\check{P}}=[\frac{2%
}{\mathbf{\check{T}}_{0}}\frac{\widehat{\mathbf{f}}_{1}^{(1)}}{\widehat{%
\mathbf{F}}^{(2)}}-\frac{\mathbf{\check{P}}_{0}}{\mathbf{\check{T}}_{0}}%
\frac{\widehat{\mathbf{F}}^{(1)}}{\widehat{\mathbf{F}}^{(2)}}\left( 2\ \ ^{m}%
\widehat{\mathcal{L}}-\mathbf{\check{T}}_{0}\right) ]\ \delta \mathbf{\check{%
R}}.$$
 The values of type $\delta \widehat{\mathbf{P}}$ and $\delta \widehat{%
\mathbf{R}}$ are considered for a small perturbations in the curvature where
$\widehat{\mathbf{R}}=\widehat{\mathbf{R}}_{0}+\delta \widehat{\mathbf{R}}$
and $\widehat{\mathbf{P}}=\widehat{\mathbf{P}}_{0}+\delta \widehat{\mathbf{P}%
}.$ No instability appears if $\delta \mathbf{\check{P}=}\delta \mathbf{%
\check{R}=0}$ which is a particular solution of the above equation. We can
in fact model a damped oscillator with additional nonholonomic constraints
if $\mathbf{\Xi }_{0}+2\mathbf{\check{P}}_{0}\widehat{\mathbf{f}}_{1}^{(1)}/%
\widehat{\mathbf{F}}^{(2)}\geq \mathbf{\check{T}}_{0},$ which allows to
avoid large instabilities in the interior of a spherical body. For some
specific functionals $f(R)$ and appropriate $G(T),$ the same behavior as in
GR results (with mass stability in the cosmological context), although there
are possible large perturbations $\delta R$ and $\delta P$ remaining. The
ideas how to circumvent the mass instability problem for holonomic
configurations has been studied in \cite{massinst}. Redefining the
generating functions and sources in a $f$-modified model into an effective
Einsteinian theory, with $\mathcal{S}[\mathbf{\check{R},}\check{\Lambda}],$
one can consider a nonholonomically deformed Hilbert-Einstein action with $%
\widehat{\mathbf{f}}\rightarrow \mathbf{\check{f}=\check{R}.}$ In such
cases, $\delta \widehat{\mathbf{R}}=\delta \mathbf{\check{R}=0}$ and
instabilities are not produced, indeed, if we impose the zero torsion
conditions (see (\ref{dtors})), we get back to the GR theory. Even if Eq.~(%
\ref{lineardef}) involves not only perturbations of the Ricci scalar $%
\widehat{\mathbf{R}}$ but also of the Ricci d-tensor $\widehat{\mathbf{R}}%
_{\alpha \beta }$ (through $\delta \widehat{\mathbf{P}}$), via nonholonomic
transforms to effective $\mathbf{\check{R}}_{\ \beta }^{\alpha }=\check{%
\Lambda}\delta _{\ \beta }^{\alpha },$ the stability of the system is
obtained via off-diagonal interactions and the nonholonomic constraints used
for an effective modeling of a subclass of $\widehat{\mathbf{f}}$-theories
to certain nonholonomic deformations of the Einstein equations with
effective cosmological constant $\check{\Lambda}.$ This is indeed a
remarkable result.

\section{Concluding remarks and discussion}

We have presented a further study of coditions when  a wide class of $%
f(R,...)$ modified gravity theories, MGTs, can be encoded into effective
off-diagonal Einstein spaces if nonholonomic deformations and constraints
are considered for the nonlinear dynamics of gravity and matter fields. A
special attention has been paid to a new version of MGT  which includes
strong coupling of the fields \cite{odgom}. Notably, we advocated such
theories to have physical motivations from the covariant Ho\v{r}ava-Lifshitz
like gravity models, with dynamical breaking of the Lorentz invariance \cite%
{covhl}. This provides also an example of a covariant, power-counting
renormalizable theory and is represented by a simplest power-law $f$%
--gravity.

It is worthwhile to mention that the gravitational field equations in such
MGTs admit a decoupling property with respect to certain classes of
nonholonomic frames, which allows us to generate exact solutions for very
general off-diagonal forms. The corresponding integral varieties of
solutions are parameterized by generating and integration functions and
various classes of commutative and noncommutative symmetry parameters. It is
possible to re-define the generating functions and effective sources of
matter fields in such a way that the $f$-terms are equivalently encoded into
effective Einstein spaces with complex parametric nonlinear structure for
the gravitational vacuum. We argue that certain nonholonomic configurations
model also covariant gravity theories with nice ultraviolet behaviors and
seem to be (super-)renormalizable in the sense of Ho\v{r}ava-Lifshitz
gravity \cite{vhl,covhl,voffdmgt}.

Notwithstanding the fact that the various $f(R)$ modified theories and
general relativity, GR,  are actually very different theories, the
off-diagonal configurations and nonlinear parametric interactions considered
in GR may encode various classes of such modified gravity effects and
explain alternatively observational data for accelerating cosmology and
certain effects in dark energy and dark matter physics. In both cases, it is
possible to find cosmological solutions and reconstruct the corresponding
action. In the already mentioned classes of modified gravity theories with $f
$-modifications \cite{revfmod,covhl,odgom}, the dynamics of the matter
sector is modeled by a perfect fluid.

We note, finally, that MGTs in general contain ghosts, due to the
higher-derivative terms in the action. However, we can select certain
ghost-free configurations determined by corresponding classes of
nonholonomic deformations or constraints. Such models of bi-metric and
massive graviton gravities were recently studied in \cite{massgr,voffdmgt}.
Together with the results in \cite{covhl,vhl}, the conclusion is reached
that some $f(R,T,R_{\alpha \beta }T^{\alpha \beta })$ models, and their
off-diagonal nonholonomic equivalents, may possess nice ultraviolet
properties and that interesting connections can be established with viable
theories of quantum gravity.

\vskip5pt

\textbf{Acknowledgments: } This work has been partially supported by the
Program IDEI, PN-II-ID-PCE-2011-3-0256, by an associated visiting research
position at CERN, by MINECO (Spain), grant PR2011-0128 and project
FIS2010-15640, by the CPAN Consolider Ingenio Project, and by AGAUR
(Generalitat de Ca\-ta\-lu\-nya), contract 2009SGR-994. We thank S.
Capozziello, S. D. Odintsov, S. Rajpoot, E. Saridakis, D. Singleton, and P.
Stavrinos for important discussions and support.

\end{document}